\documentclass[preprint]{aastex}

\def\bfc{}

\usepackage{graphicx}
\usepackage{epsf}
\usepackage{natbib}

\newcommand{\simlt}{\lower.5ex\hbox{$\; \buildrel < \over \sim \;$}}

\providecommand{\sorthelp}[1]{}

\slugcomment{draft date: \today}

\begin{document}

\title{Variations between Dust and Gas in the Diffuse Interstellar Medium}

\shorttitle{Variations between Dust and Gas}

\author{William T. Reach}
\affil{Universities Space Research Association, MS 232-11, Moffett Field, CA 94035, USA}
\email{wreach@sofia.usra.edu}
\and
\author{Carl Heiles}
\affil{Astronomy Department, University of California, Berkeley, CA 94720, USA}
\and
\author{Jean-Philippe Bernard}
\affil{Universit\'e de Toulouse, Institut de Recherche en Astrophysique et Plan\'etologie, F-31028 Toulouse cedex 4, France}

\begin{abstract}
Using the {\it Planck} far-infrared and Arecibo GALFA 21-cm line surveys, we identified a set of isolated interstellar clouds (approximately degree-sized on
the sky and comprising 100 solar masses) and assessed the ratio of gas mass to dust mass. 
Significant variations of the gas-to-dust ratio are found both from cloud to cloud and within regions of individual clouds; within the clouds, the atomic gas per unit dust 
decreases by more than a factor of 3 compared to the standard gas-to-dust ratio. 
Three hypotheses are considered. First, the apparently {\bfc low} gas-to-dust ratio could be due to molecular gas. Comparing to {\it Planck}\ CO maps, the brightest clouds have a 
H$_2$/CO ratio comparable to galactic plane clouds, but a strong lower limit is placed on the ratio for other clouds, such that the required amount of molecular gas is far higher than would be expected based on the CO upper limits. Second, we consider self-absorbed 21-cm lines and find the optical depth must be $\sim3$, significantly higher than found from surveys of radio sources. Third, grain properties may change within the clouds: they become more emissive when they are colder, while not utilizing heavy elements that already have their cosmic abundance fully locked into grains. It is possible all three processes are active, and follow-up studies will be required to disentangle them and measure
the true total gas and dust content of interstellar clouds.
\end{abstract}


\section{Introduction}

The nature of the diffuse interstellar medium, which pervades the volume of the galaxy and is
distinct from the dense, self-gravitating clouds where stars form, is only understood in broad
terms. Studies of absorption toward nearby stars reveal a combination of atomic and molecular gas
toward those lines of sight \citep{savage77,radford02}, but they leave large gaps between the observed stars and have relatively
little context by which we can understand the organization of the material or its relation to material
not readily studied by absorption lines. Large-scale surveys of 
emission from diffuse interstellar gas were made possible by the brightness of the 21-cm hyperfine
line of H~I and dedicated all-sky surveys at half-degree angular resolution
\citep{kalberla05}. 
Widespread emission from interstellar dust was revealed by the first infrared satellite observations and
unveiled even more detail about the organization of the interstellar medium \citep{low84}. The structure of the interstellar clouds and the relation  between gas and dust could be studied
at half-degree resolution, limited by the 21-cm surveys. 
Dust and gas well correlated at that scale;
however, the infrared images  showed
significant structure on smaller scales, and variations between dust and gas appeared significant
on small scales in limited observations performed. As one observer noted,
``[o]ur results emphasize that it is not always evident, from observations with spatial resolutions coarser than a few arc minutes, how interstellar gas and dust are related'' \citep{verschuur92}.

At the level of individual clouds, where a `cloud' is defined as a structure on the sky of high 
spatial contrast on the $<1^\circ$ scale as well as a velocity coherent at the $<10$~km~s$^{-1}$ scale,
there are significant variations between the dust and gas properties. The advantage of working on 
features of relatively high contrast is that we can isolate the structures by subtracting the brightness
just outside the cloud, allowing measurements of the cloud without requiring knowledge of the 
absolute goal was
 brightness. For degree-sized clouds selected from the {\it Infrared Astronomical
Satellite} ({\it IRAS}), we found significant variations in the relative amounts of dust and gas from
cloud to cloud \citep{hrk88} despite the overall linear correlation between dust and gas when averaged over very large areas \citep{boulangerperault88}. 
In particular, there was more dust emission per unit H~I column density for some
clouds, suggesting the dust was tracing additional material not traced by the 21-cm line observations. 
Follow-up observations showed CO in several of those clouds {\bfc indicated the extra
infrared emission was often associated with molecular gas, though the correspondence was
not perfect \citep{reach94,meyerdierks96}. Some galactic CO emission is apparently present in small clumps,
even at high galactic latitude \citep{heithausen02}.}

All-sky comparisons of dust and gas emission, include far-infrared observations out to 240 $\mu$m wavelength with the {\it Cosmic Background Explorer} at degree scales 
showed that locations with apparent `excess' infrared
emission tended to also be {\it colder} than their surroundings (so the excess is truly in the
amount of dust, not the amount of heating) and provided an all-sky guide to locations
of potential molecular clouds \citep{reach98}. 
More recently, the all-sky survey by {\it Planck} with its exceptionally high sensitivity {\bfc at $5'$
angular scales, confirmed the presence and relative frigidity of the infrared excess and 
showed that the amount of material that is implicated in producing the excess is comparable to the
known molecular mass} of the local interstellar medium \citep{planck2011-7.0}.

The potential for molecular gas without CO has been noted for some time \citep{blitz90}, but the true amount and extent of such gas is only recently
becoming elucidated by observations such as mentioned above and described in this paper. 
H$_2$ is self-shielded and forms in regions protected front he interstellar radiation field by approximately half a magnitude of extinction, while CO requires 
somewhat more protection at approximately 2 magnitudes of extinction. 
A theoretical prediction for the amount of CO-dark 
molecular gas yields results that can generally agree with those inferred from observations, with a reasonable set of
physical properties of the interstellar medium, depending primarily on the fraction of the total medium that is in very large clouds \citep{wolfire10}. 
Therefore, in regions with no giant molecular clouds, the fraction of `dark' molecular gas is expected to be high.
The potential existence of untraced molecular gas is important for understanding the formation of molecular clouds on galactic scales, where large-scale
molecular (but low-density) filaments are predicted to extend well beyond the observed dense filaments \citep{smith14}.

In this paper, we study isolated structures in the local interstellar medium in order to compare
dust and gas at $5'$ angular scales. We address the questions of whether dust and gas
are well mixed, what is the relative amount of atomic and molecular gas in the interstellar medium,
and where does the gas transform from atomic to molecular.

\section{Observations\label{obssec}}

\subsection{Arecibo}
\def\hi{H{\small{I}}}
The 21-cm line data used in this work were part of the large-scale Galactic Arecibo L-band Feed Array (GALFA) \hi\ survey, Data Release 1 \citep{peek11}. The observations with the 305-m Arecibo telescope are part of a survey that will cover the entire sky visible from that telescope (declination $0^\circ$ to $40^\circ$) with an angular resolution of $4'$. The time-ordered spectra were regridded into cubes and presented in an online archive. We extracted cubes centered on distinct interstellar clouds evident in the {\it Planck} all-sky survey. 
An interference spike was removed from the cube near G94-36, and linear baselines were refit using symmetrically located portions on either side of the low-velocity H~I that dominates the spectra. 
For each position, we integrate the 21-cm line and convert to a column density assuming that the line is optically thin. We return to address this assumption later in the paper.
Note that the Arecibo observations contain stray-light due to 21-cm emission from distant sources (in particular, the galactic plane) entering the far side lobes of the beams of the 7 receivers. 
{\bfc 
Comparison with the stray-light-corrected all-sky survey at low angular resolution \citep{kalberla05} show the amplitude of stray light is less than $\sim 1$ K brightness temperature, which is much less than the brightness of the clouds studied in this paper ($>10$ K).
}
The stray-light should be smooth on the $\sim$degree angular scale and smaller that are of interest in this project. The stray-light should also have a relatively broad velocity profile (because it contains large portions of the rotating galactic plane) which we verify is not evident at the levels $>10^{20}$ cm$^{-2}$ of interest to this project.

\subsection{{\it Planck}}

The {\it Planck} survey covered the entire sky in 9 bands with wavelengths from 30 GHz to 857 GHz (1 cm to 350 $\mu$m)
\citep{tauber2010a, Planck2011-1.1}. 
The wide range of frequencies allowed a separation of the sky into the cosmic microwave background and astrophysical foregrounds \citep{planck2013-p06}; we utilized the public release of the `thermal dust' component. The optical depth of interstellar dust at 353 GHz was derived from a fit to the combination of the {\it Planck} 353, 545, and 857 GHz (wavelengths 850, 550, and 350 $\mu$m) from the
High-Frequency Instrument \citep{planck2013-p03f}
together with the {\it IRAS} 100 $\mu$m data as reprocessed to combine {\it COBE}/DIRBE large-scale calibration \citep{miville05}. 
After the contaminating signals from the CMB and zodiacal light, the dust emission was fitted with
\begin{equation}
\label{eq:inu}
I_\nu = \tau(353\,{\rm GHz}) \left(\frac{\nu}{353\,{\rm GHz}}\right)^\beta B_\nu(T_d)
\end{equation}
where $\tau$ is the optical depth and $\beta$ is an emissivity power-law index that approximates the 
frequency-dependence of the dust emissivity. 
The fit for $\beta$ was performed at $30'$ resolution, and then used to make a finer-scaled fit for $T_d$ at
the full $5'$ resolution.

\subsection{Cloud selection}

\begin{figure*}[tbh]
\includegraphics[scale=0.32]{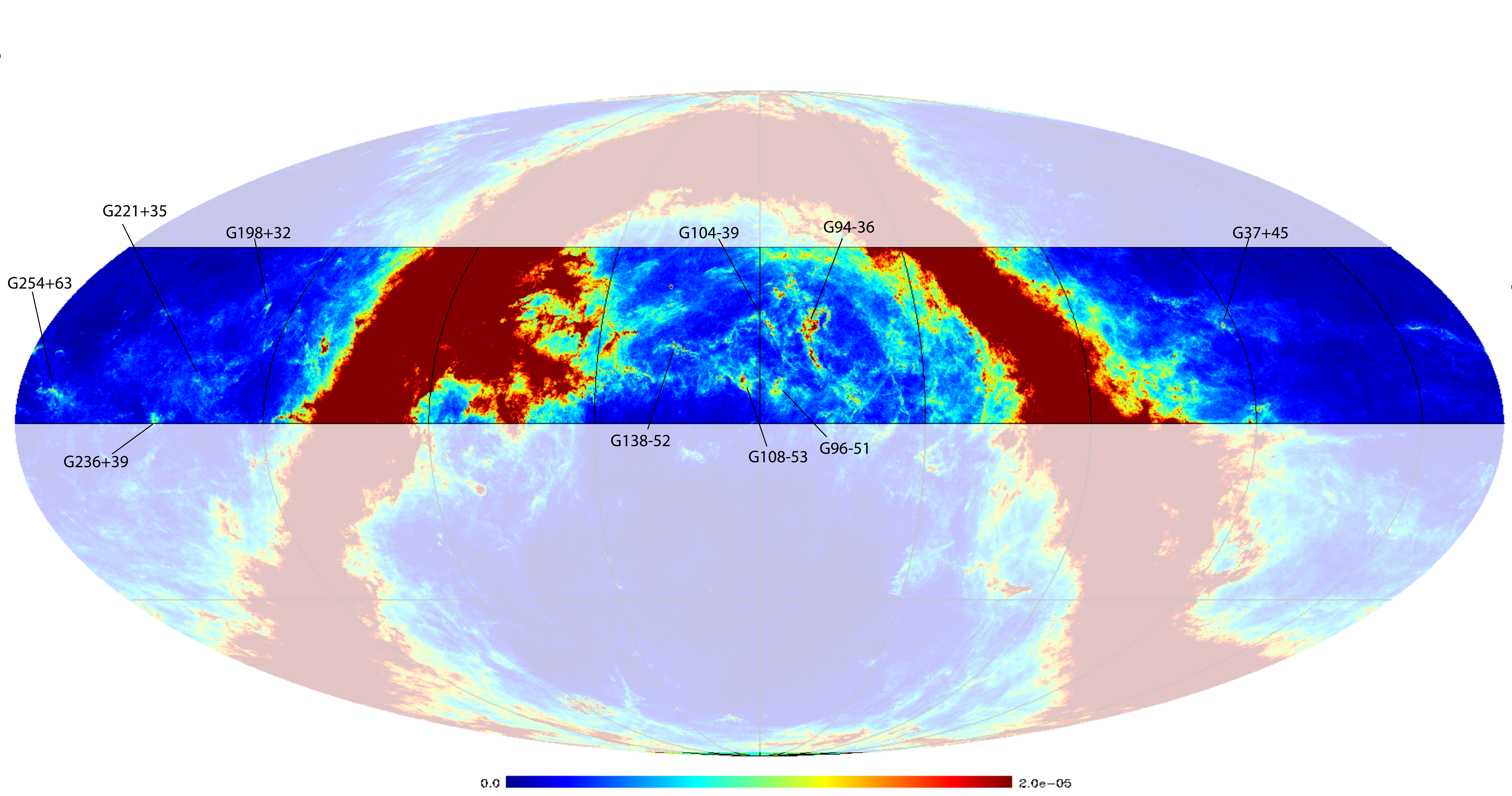}
\caption[tauEQlab]{
All-sky map of the {\it Planck} 353 GHz optical depth, $\tau({\rm 353\, GHz)}$, in celestial coordinates. The declination range not visible from Arecibo is greyed out. Each of the 
relatively isolated interstellar clouds addressed in this study are labeled with their name (based on galactic coordinates). 
\label{tauEQlab}}
\end{figure*}

Our goal was to identify isolated, degree-sized clouds for which the unrelated background ISM could be separated from a distinct spatially compact (at $5'$ resolution) structure.
The clouds studied in this paper were selected by inspecting the {\it Planck} 353 GHz map, in equatorial coordinates, within the Arecibo declination range $-1^\circ < \delta < 38^\circ$ and clearly separated from the ridge of emission centered in the galactic plane. Figure~\ref{tauEQlab} shows the clouds on the {\it Planck} image, and Table~\ref{obstab} lists their locations.
Despite being the prominent, isolated features in the Arecibo-visible portion of the sky, most of the clouds have not been previously studied.
G94-36 includes MBM 53 and 54, which had been identified in a survey of the Palomar plates for visible extinction
\citep{mbmfirst}.
G104-39 includes DIR\,105-38 , an infrared excess cloud identified from comparison of {\it Cosmic Background
Explorer}/Diffuse Background Experiment (DIRBE) far-infrared observations to large-scale 21-cm survey \citep{reach98}.

\def\tnm{\tablenotemark}
\begin{table}
\caption[]{Properties of the Interstellar Clouds in this Study\tnm{a}}\label{obstab} 
\begin{flushleft} 
\begin{tabular}{llrcccc}
\hline\hline
Name &  Coordinates & $v_0$ & $\delta V$ & $T_d$ & $\beta$  & $\Delta N({\rm H~I})$\\ 
           & (J2000) & (km~s$^{-1}$) &  (km~s$^{-1}$) & (K) &         & (10$^{20}$ cm$^{-2}$)\\ 
\hline
G94-36     & 23:05:23 +20:59 &  -7.0 & $4.90\pm 0.08$  & $17.0\pm 0.3$ & $1.78\pm 0.01$ & 4.2  \\
G96-51     & 23:43:14 +08:58 &  -9.1 & $6.61\pm 0.11$  & $18.0\pm 0.3$ & $1.64\pm 0.02$ & 4.3 \\
G104-39    & 23:48:42 +21:50 & -13.6 & $6.8\pm 0.2 $   & $17.4\pm 0.3$ & $1.79\pm 0.02$ & 3.0 \\
G108-53    & 00:17:06 +09:45 &  -5.7 & $8.70\pm 0.14$  & $17.0\pm 0.2$ & $1.84\pm 0.01$ & 4.6    \\
G138-52    & 01:29:26 +10:10 &  -9.5 & $6.45\pm 0.05$  & $18.7\pm 0.2$ & $1.71\pm 0.01$ & 4.4  \\ 
G198+32    & 08:27:26 +26:12 &  +5.1 & $2.91\pm 0.02$  & $18.4\pm 0.7$ & $1.45\pm 0.03$ & 3.0 \\ 
G198+32 E\tnm{b} & 08:16:15 +27:02 &  +5.1 & $2.9\pm 0.02$   & $19.3\pm 0.3$ & $1.70\pm 0.02$ & 1.5 \\
G221+35    & 09:08:42 +09:26 &  +6.1 & $10.1\pm 0.2$   & $18.6\pm 0.2$ & $1.66\pm 0.01$  & 5.0 \\
G236+39  & 09:45:58 +00:35 & +10.2 & $8.1\pm 0.5$  & $18.0\pm 0.3$ & $1.64\pm 0.02$ & 3.0 \\
G236+39 N\tnm{c}  & 09:46:46 +01:51 & +9.3 & $5.0\pm 0.2$  & $18.9\pm 0.2$ & $1.73\pm 0.02$ & 5.7 \\
G254+63    & 11:29:18 +07:34 &  -1.6 & $6.22\pm 0.09$  & $18.9\pm 0.4$ & $1.25\pm 0.01$ & 4.7 \\
\end{tabular}
\tablenotetext{a}{$T_d$ and $\beta$ are the average dust temperature (K) and emissivity index over the cloud peaks, and the numbers after the $\pm$ are standard deviations within the cloud. {\bfc $\Delta N({\rm H~I})$ is the 21-cm column density (in the optically-thin limit) toward the cloud peak minus a background within the 8$^\circ$ region containing the cloud.}
}
\tablenotetext{b}{cloud near G198+32 with high gas/dust}
\tablenotetext{c}{northern component of G236+39}
\end{flushleft} 
\end{table}  

\begin{table}
\caption[]{Derived properties\tnm{a}}\label{derivtab} 
\begin{flushleft} 
\begin{tabular}{lccccccc}
\hline\hline
Name &  $T_{\rm diff}$ & $\left[G/D\right]_{\rm diff}$  &  
$T_{\rm cloud}$ & $\left[ G/D \right]_{\rm cloud}$ & $f_{\rm dark}$ & Cloud Mass & H~I Mass \\ 
 & (K) & & (K) & & & ($M_\odot)$ & ($M_\odot$) \\
\hline
G94-36  & $18.5\pm 0.6$& $48 \pm 11 $& $17.1\pm 0.2$ & $17\pm 2$ & 2.9 & 106 & 22 \\ 
G96-51  & $18.3\pm 0.4$& $51 \pm 3 $& $17.5\pm 0.2$ & $24\pm 3$ & 2.1 & 72 & 17 \\ 
G104-39 & $18.4\pm 0.5$& $53 \pm 9 $ & $17.4\pm 0.2$ & $20\pm 2$ & 2.6 & 91 & 14 \\ 
G108-53 & $18.1\pm 0.5$& $38 \pm 3 $& $16.5\pm 0.2$ & $8.7\pm 2$ & 4.4 & 140 & 35 \\ 
G138-52 & $18.5\pm 0.5$& $80 \pm 6 $& $18.6\pm 0.2$ & $80\pm 6$ & 1.0 & 52 & 38 \\ 
G198+32 & $20.0\pm 0.5$& $53 \pm 9 $& $18.4\pm 0.2$ & $16\pm 2$ & 3.2 & 60 & 12 \\ 
G198+32 E   & $20.0\pm 0.5$& $80 \pm 29 $& $20.0\pm 0.2$ & $138\pm 19$ & 0.6 & 7 & 7 \\ 
G221+35 & $19.1\pm 0.4$& $97 \pm 5$& $18.3\pm 0.2$ & $48\pm 3$ & 2.0 & 32 & 22 \\ 
G236+39 & $19.5\pm 0.6$& $74 \pm 17 $& $18.4\pm 0.2$ & $42\pm 2$ & 1.8 & 45 & 30 \\ 
G236+39 N & $19.5\pm 0.6$& $51 \pm 5 $& $19.0\pm 0.5$ & $54\pm 6$ & 1.0 & 28 & 22 \\ 
G254+63 & $20.5\pm 0.8$& $121 \pm 14 $& $18.9\pm 0.2$ & $33\pm 2$ & 3.6 & 84 & 27 \\ 
\end{tabular}
\tablenotetext{a}{$\left[ G/D \right]_{\rm cloud}$ and $\left[ G/D \right]_{\rm diff}$ are gas-to-dust mass ratio inferred from the slope of dust optical depth versus H~I column 
density for the clouds and for the regions just outside the clouds, respectively. For reference the average interstellar medium value is 124 \citep{lidraine}.
The `dark' factor, $f_{\rm dark}$ is defined in Eq.~\ref{eq:fdark}.}
\end{flushleft} 
\end{table}

\clearpage

\begin{figure}[th]
\includegraphics[scale=.7]{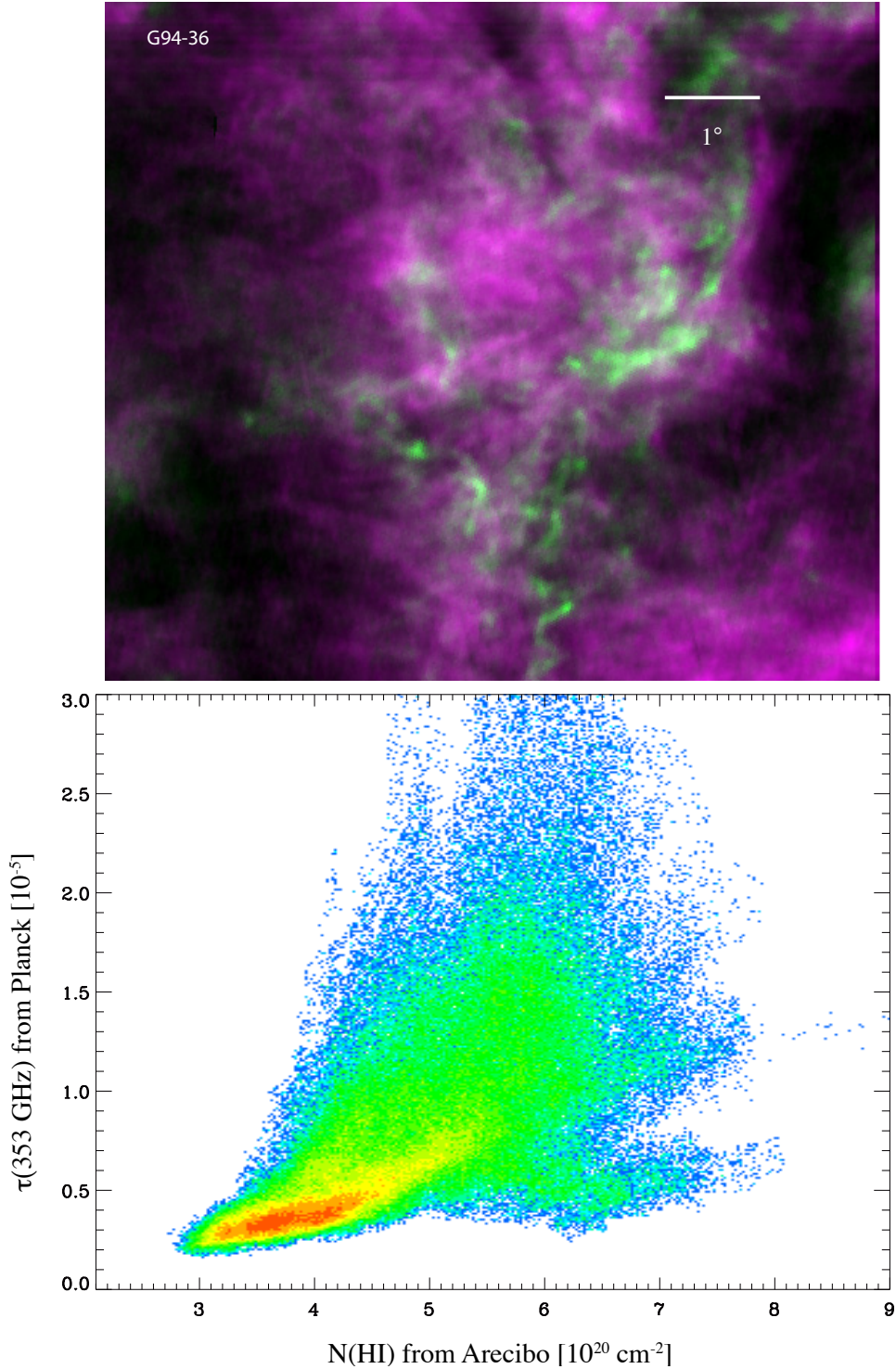}
\caption[G94mainFigure]{
{\it(Top panel)} An image covering $8.2^\circ\times 7.5^\circ$ centered on J2000 coordinates 23:11:51 +21:11 (galactic coordinates 93.9, -36.0) with celestial North upward. The color composite was constructed with the gas column density (inferred in the optically-thin limit) from the Arecibo 21-cm line in red+blue and the dust column density inferred from the {\it Planck} 
optical depth in green. 
{\it(Bottom panel)} Hess diagram (densitized scatter plot) of the dust versus gas column density using the pixel values from the image shown in the top panel. The density of points in the scatter plot is shown in a continuous color table from red (high-density) to blue (low-density). 
\label{G94mainFigure}}
\end{figure}

\begin{figure}[th]
\includegraphics[scale=.7]{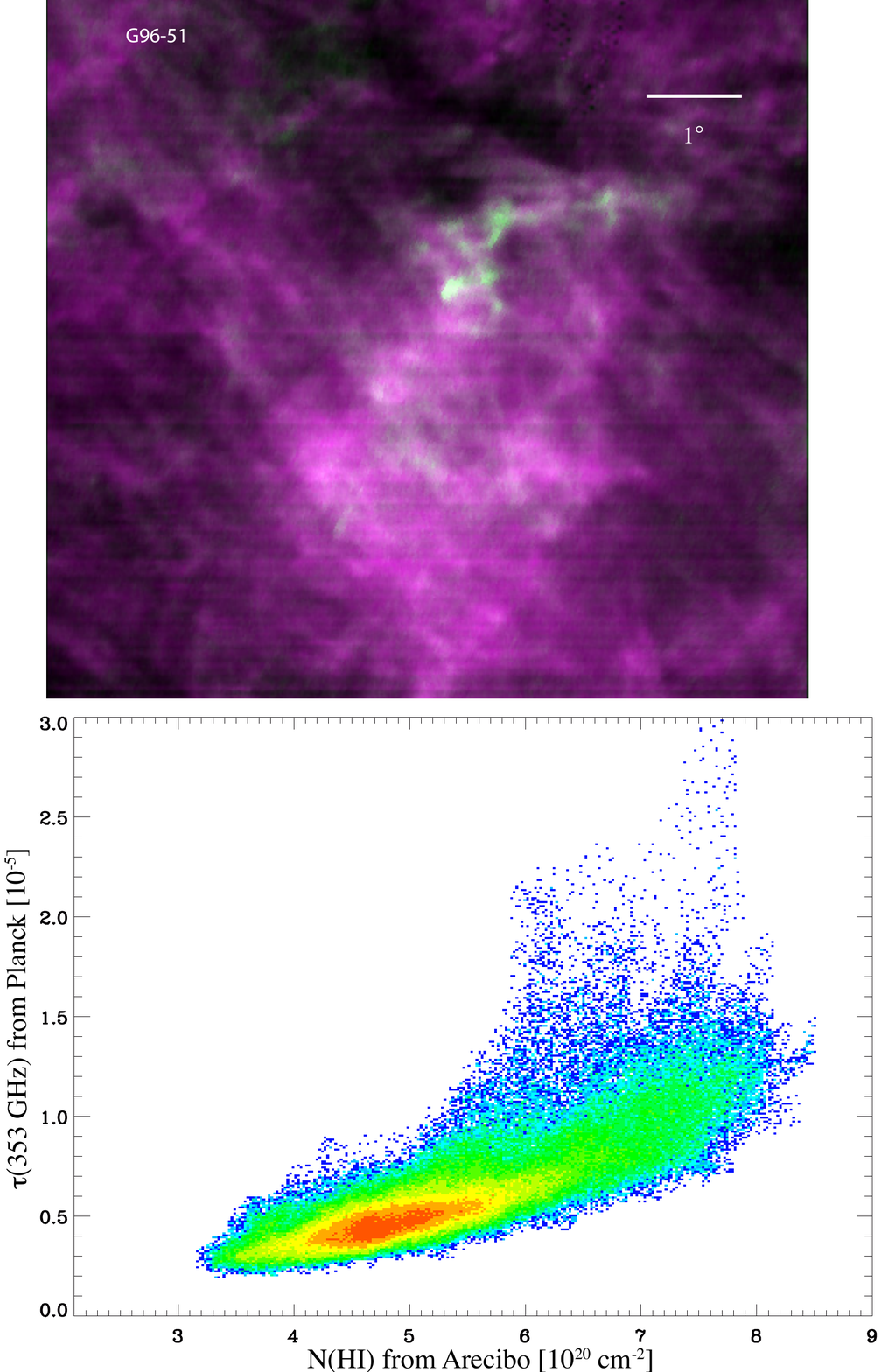}
\caption[G96mainFigure]{
{\it(Top panel)} An image covering $8.2^\circ\times 7.4^\circ$ centered on J2000 coordinates 23:44:14 +08:17 (galactic coordinates 95.9, -51.0) with celestial North upward. The color composite was constructed with the gas column density (inferred in the optically-thin limit) from the Arecibo 21-cm line in red+blue and the dust column density inferred from the {\it Planck} 
optical depth in green. 
{\it(Bottom panel)} Hess diagram (densitized scatter plot) of the dust versus gas column density using the pixel values from the image shown in the top panel. The density of points in the scatter plot is shown in a continuous color table from red (high-density) to blue (low-density). 
\label{G96mainFigure}}
\end{figure}

\begin{figure}[th]
\includegraphics[scale=.7]{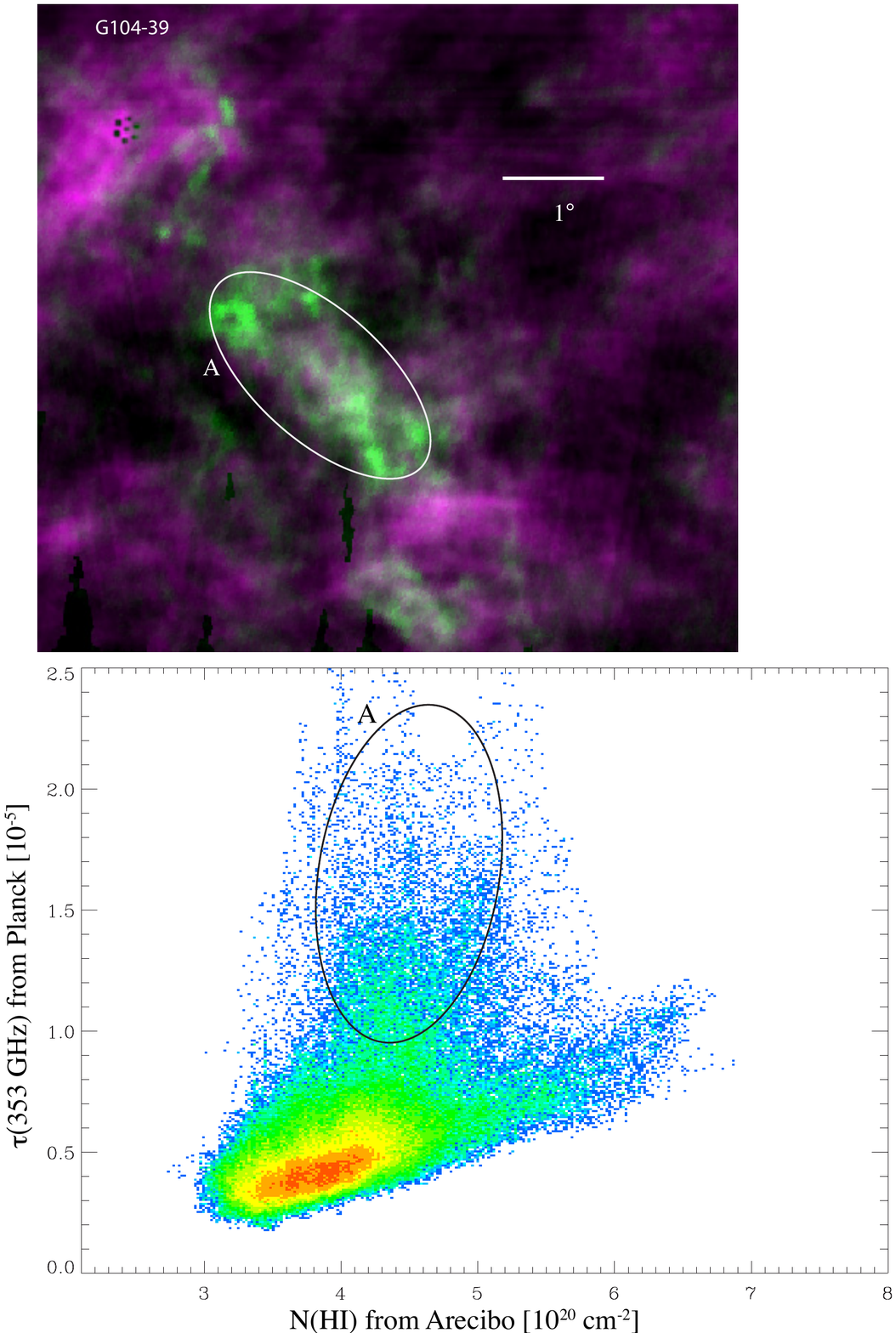}
\caption[G04mainFigure]{
{\it(Top panel)} An image covering $7.1^\circ\times 6.4^\circ$ centered on J2000 coordinates 23:46:35 +22:17 (galactic coordinates 103.7, -38.2) with celestial North upward. The color composite was constructed with the gas column density (inferred in the optically-thin limit) from the Arecibo 21-cm line in red+blue and the dust column density inferred from the {\it Planck} 
optical depth in green. 
{\it(Bottom panel)} Hess diagram (densitized scatter plot) of the dust versus gas column density using the pixel values from the image shown in the top panel. The density of points in the scatter plot is shown in a continuous color table from red (high-density) to blue (low-density). 
\label{G104mainFigure}}
\end{figure}

\begin{figure}[th]
\includegraphics[scale=.7]{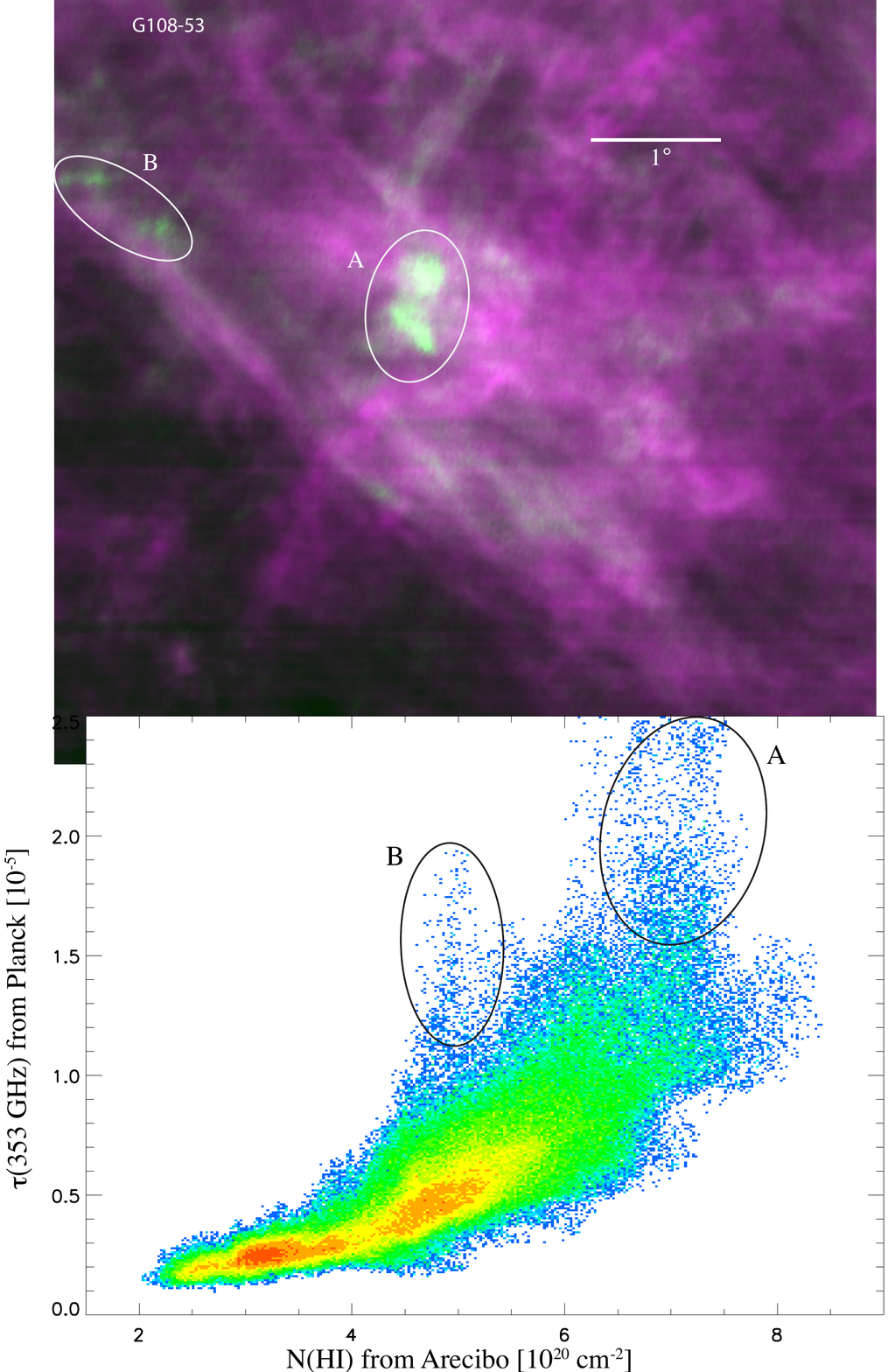}
\caption[G108mainFigure]{
{\it(Top panel)} An image covering $8.0^\circ\times 7.4^\circ$ centered on J2000 coordinates 00:15:09 +08:53 (galactic coordinates 108.0, -52.9) with celestial North upward. The color composite was constructed with the gas column density (inferred in the optically-thin limit) from the Arecibo 21-cm line in red+blue and the dust column density inferred from the {\it Planck} 
optical depth in green. 
{\it(Bottom panel)} Hess diagram (densitized scatter plot) of the dust versus gas column density using the pixel values from the image shown in the top panel. The density of points in the scatter plot is shown in a continuous color table from red (high-density) to blue (low-density). 
\label{G108mainFigure}}
\end{figure}

\begin{figure}[th]
\includegraphics[scale=.7]{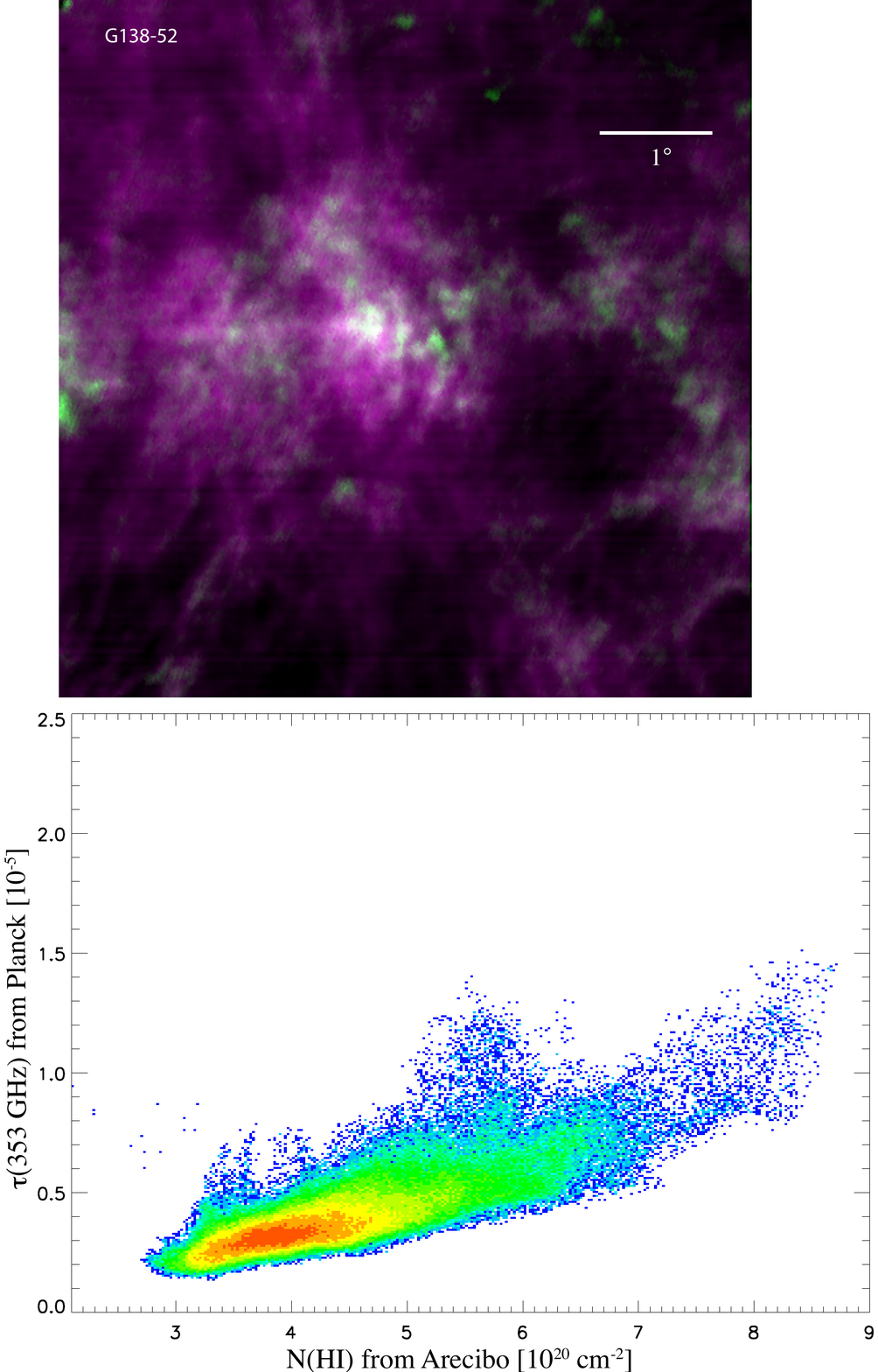}
\caption[G138mainFigure]{
{\it(Top panel)} An image covering $7.1^\circ\times 6.4^\circ$ centered on J2000 coordinates 01:27:51 +09:40 (galactic coordinates 137.7, -52.2) with celestial North upward. The color composite was constructed with the gas column density (inferred in the optically-thin limit) from the Arecibo 21-cm line in red+blue and the dust column density inferred from the {\it Planck} 
optical depth in green. 
{\it(Bottom panel)} Hess diagram (densitized scatter plot) of the dust versus gas column density using the pixel values from the image shown in the top panel. The density of points in the scatter plot is shown in a continuous color table from red (high-density) to blue (low-density). 
\label{G138mainFigure}}
\end{figure}

\begin{figure}[th]
\includegraphics[scale=.7]{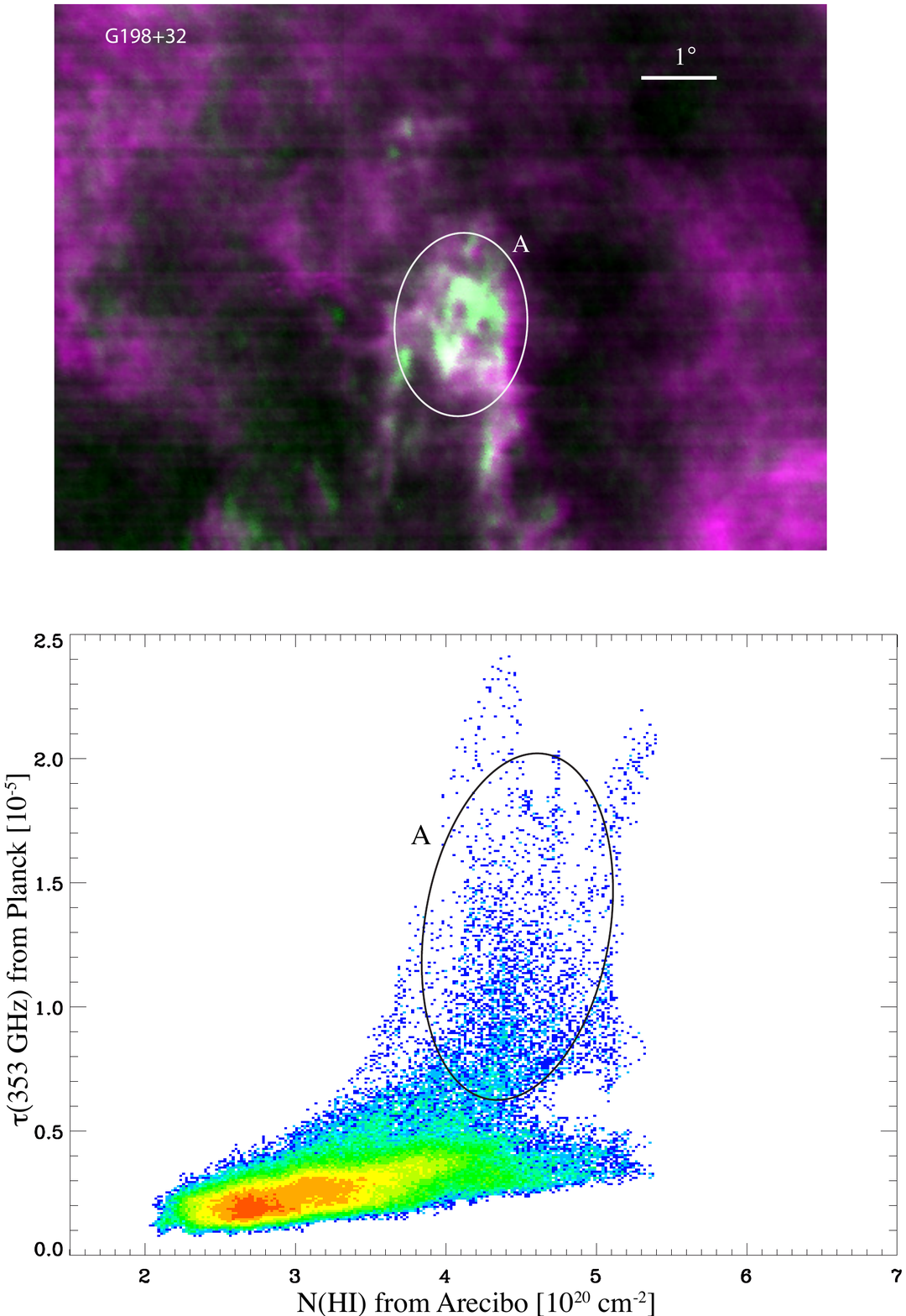}
\caption[G198mainFigure]{
{\it(Top panel)} An image covering $8.3^\circ\times 5.8^\circ$ centered on J2000 coordinates 08:29:24 +26:29 (galactic coordinates 197.3, +32.2) with celestial North upward. The color composite was constructed with the gas column density (inferred in the optically-thin limit) from the Arecibo 21-cm line in red+blue and the dust column density inferred from the {\it Planck} 
optical depth in green. 
{\it(Bottom panel)} Hess diagram (densitized scatter plot) of the dust versus gas column density using the pixel values from the image shown in the top panel. The density of points in the scatter plot is shown in a continuous color table from red (high-density) to blue (low-density). 
\label{G198mainFigure}}
\end{figure}

\begin{figure}[th]
\includegraphics[scale=.7]{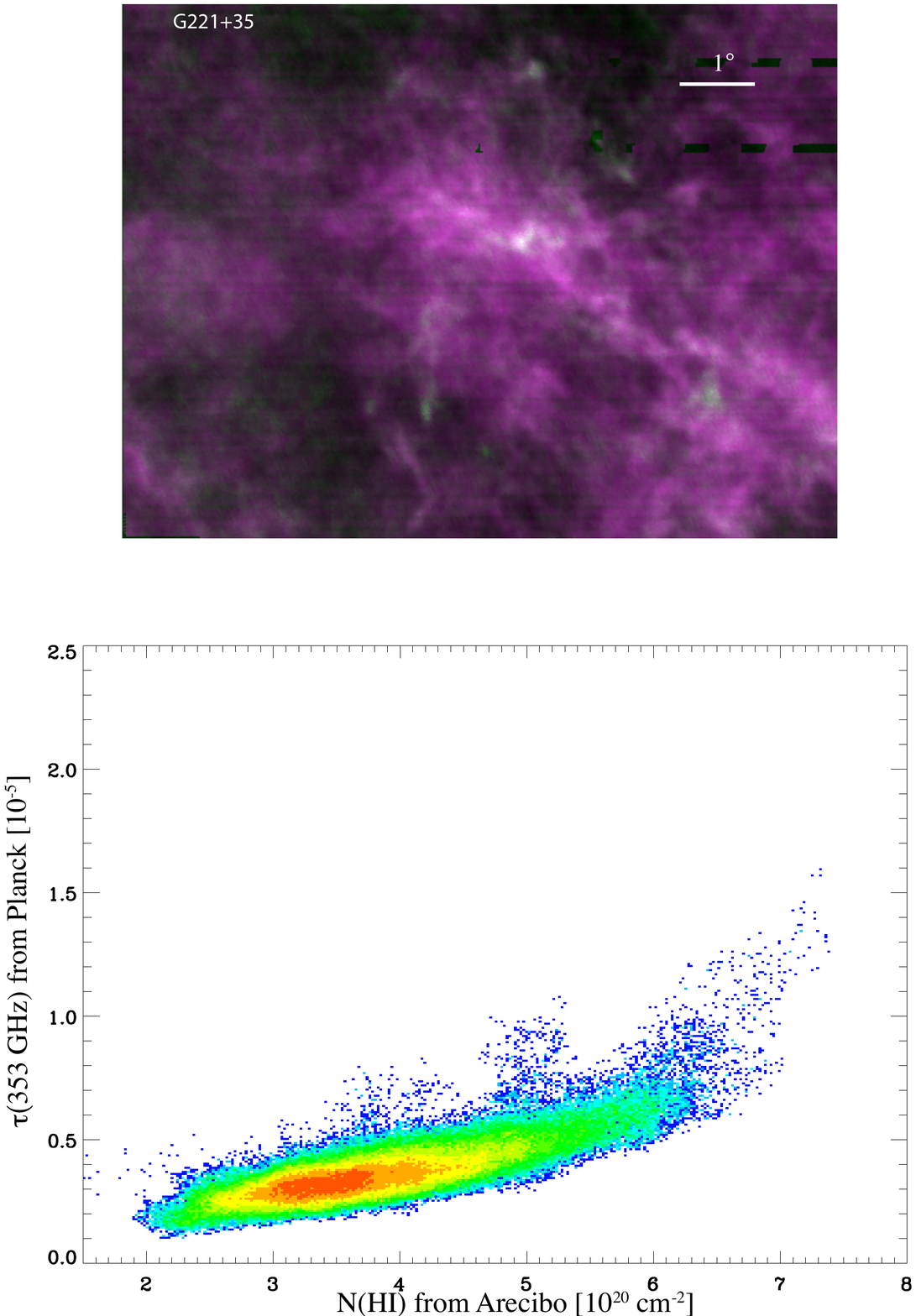}
\caption[G221mainFigure]{
{\it(Top panel)} An image covering $8.1^\circ\times 5.9^\circ$ centered on J2000 coordinates 09:10:15 +09:01 (galactic coordinates 221.1, +34.9) with celestial North upward. The color composite was constructed with the gas column density (inferred in the optically-thin limit) from the Arecibo 21-cm line in red+blue and the dust column density inferred from the {\it Planck} 
optical depth in green. 
{\it(Bottom panel)} Hess diagram (densitized scatter plot) of the dust versus gas column density using the pixel values from the image shown in the top panel. The density of points in the scatter plot is shown in a continuous color table from red (high-density) to blue (low-density). 
\label{G221mainFigure}}
\end{figure}

\begin{figure}[th]
\includegraphics[scale=.7]{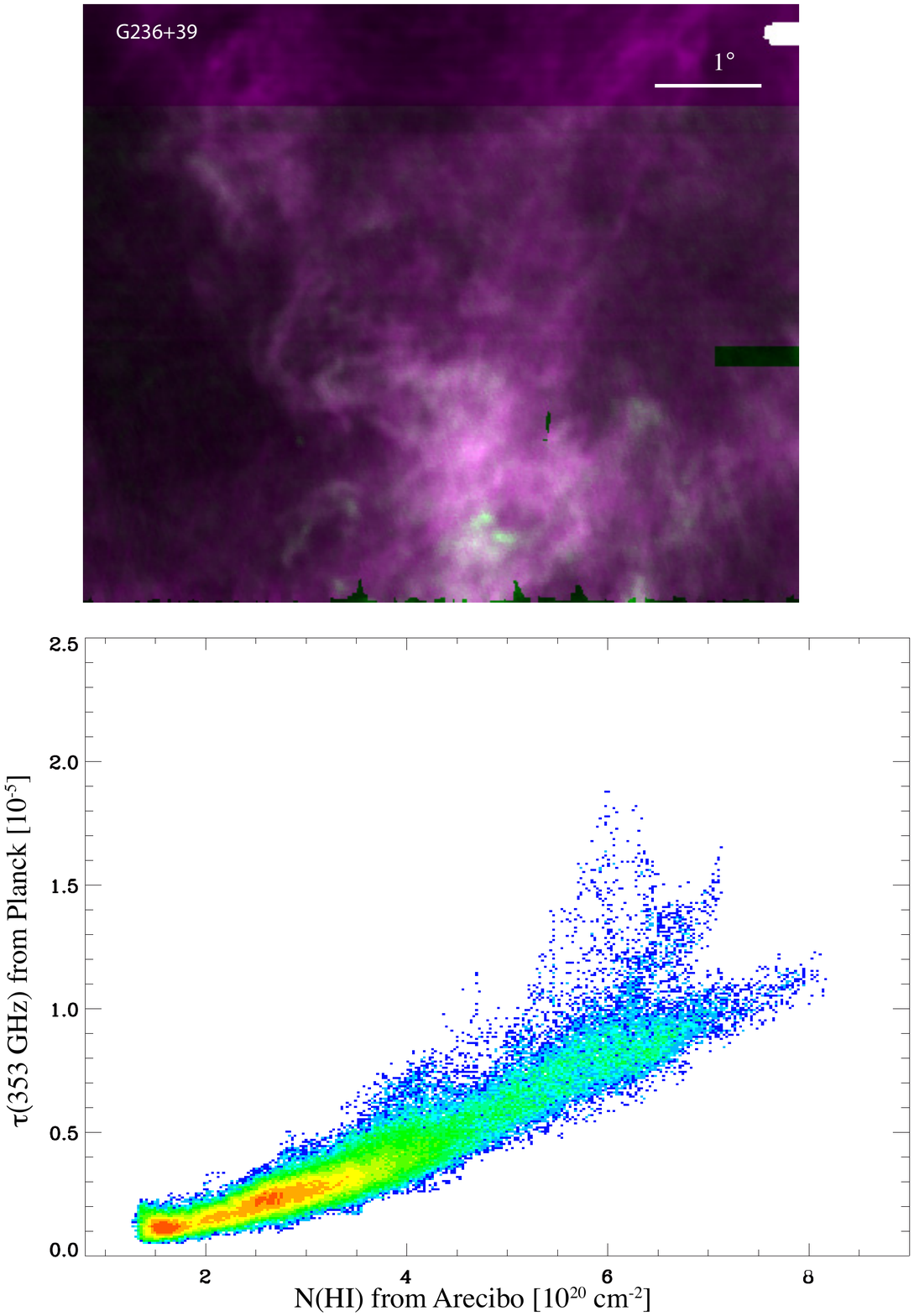}
\caption[G236mainFigure]{
{\it(Top panel)} An image covering $7.7^\circ\times 4.5^\circ$ centered on J2000 coordinates 09:47:04 +02:47 (galactic coordinates 233.7, +39.7) with celestial North upward. The color composite was constructed with the gas column density (inferred in the optically-thin limit) from the Arecibo 21-cm line in red+blue and the dust column density inferred from the {\it Planck} 
optical depth in green. 
{\it(Bottom panel)} Hess diagram (densitized scatter plot) of the dust versus gas column density using the pixel values from the image shown in the top panel. The density of points in the scatter plot is shown in a continuous color table from red (high-density) to blue (low-density). 
\label{G236mainFigure}}
\end{figure}

\begin{figure}[th]
\includegraphics[scale=0.5]{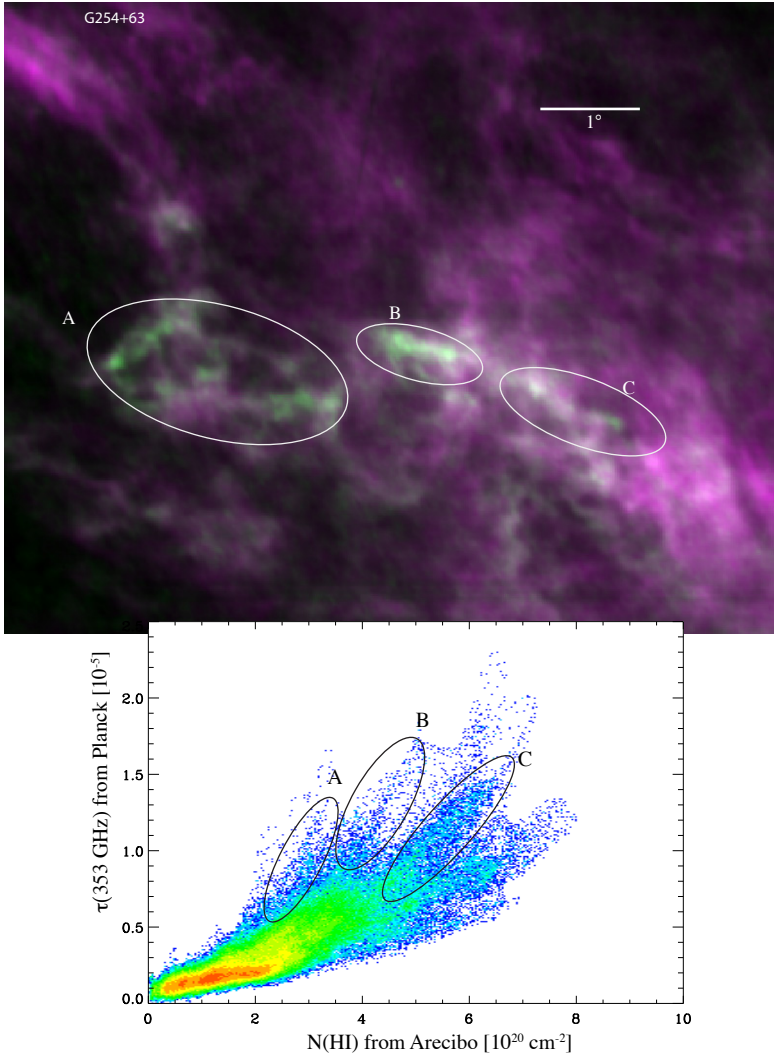}
\caption[G254mainFigure]{
{\it(Top panel)} An image covering $8.0^\circ\times 7.4^\circ$ centered on J2000 coordinates 11:30:40 +07:51 (galactic coordinates 254.4, +62.8) with celestial North upward. The color composite was constructed with the gas column density (inferred in the optically-thin limit) from the Arecibo 21-cm line in red+blue and the dust column density inferred from the {\it Planck} 
optical depth in green. 
{\it(Bottom panel)} Hess diagram (densitized scatter plot) of the dust versus gas column density using the pixel values from the image shown in the top panel. The density of points in the scatter plot is shown in a continuous color table from red (high-density) to blue (low-density). 
\label{G254mainFigure}}
\end{figure}

\clearpage

Several peaks within the clouds are listed in the {\it Planck} Galactic Cold Core Catalog (PGCC), which contains all compact cores from the {\it Planck} survey with dust temperature significantly
colder than surroundings  \citep{pgcc}. Table~\ref{pgcctab} lists the cataloged cold cores within our sample clouds together with the clump dust temperature and emissivity index. 
The weighted mean dust temperature is $\langle T_{\rm clump}\rangle=14.85\pm 0.07$ K, and the weighted mean emissivity index is 
$\langle \beta\rangle =1.81\pm 0.31$. While the $\beta$ values are comparable to those we estimated from entire clouds, the temperatures in the PGCC are colder.
This measures the decrease of dust temperature and increase of $\beta$ that continues from the diffuse medium (average $T_d=20.8$ K, $\beta=1.55$ over the faintest 10\% of the sky \citep{planckAllSkyDust}) 
to the isolated clouds ($\langle T_d\rangle=18.1\pm 0.1$, $\langle \beta\rangle=1.66$ from Table~\ref{derivtab}) down to the smaller scale of the cold cores. 
For some of the cold cores, there are distance estimates in the PGCC based on 
reddening of SDSS stars.
For G94-36, the distances are in the range 203--281 pc (and one at 421 pc);  for G198+32 the distance estimates are 290--360 pc; and
for G103-39 the distances estimate is 241 pc. The distance estimates should be treated with caution because if the low extinctions and uncertainties
in stellar distances, but the suggest at least some of the clouds are further than the fiducial 100 pc we would normally assume for high-latitude interstellar clouds
within the scale height of cold atomic gas. 

\begin{table}
\caption[]{Cold Cores within the Isolated Clouds}\label{pgcctab} 
\begin{flushleft} 
\begin{tabular}{llccc}
\hline\hline
Cloud &  PGCC\tablenotemark{a} & $T_{\rm clump}$ & $\beta_{\rm clump}$ \\ 
      & & (K) & \\
\hline
G104-39 & G105.76-38.36 & $14.8\pm 1.7$ & $1.98\pm 0.38$ \\
                 & G106.06-38.39 & $14.7\pm 2.5$ & $1.79\pm 0.53$ \\
                 & G105.07-38.06 & $15.3\pm 1.5$ & $1.96\pm 0.27$ \\
                 & G103.35-38.93 & $14.0\pm 2.4$ & $1.93\pm 0.62$ \\
                 & G103.38-39.33 & $14.9\pm 1.9$ & $2.04\pm 0.39$ \\
G108-53 & G108.74-52.67 & $15.5\pm 1.5$ & $1.78\pm 0.29$ \\
                 & G109.26-52.52 & $13.5\pm 3.6$ & $2.19\pm 1.11$ \\
G198+32 & G198.25+32.30 & $17.1\pm 3.1$ & $1.22\pm 0.61$ \\
                  & G197.22+31.99 & $15.6\pm 3.4$ & $1.40\pm 0.80$ \\
                  & G197.37+31.73 & $16.1\pm 3.6$ & $1.55\pm 0.79$ \\
                  & G197.91+31.99 & $14.9\pm 3.2$ & $1.65\pm 0.74$ \\
                  & G198.40+32.26 & $18.3\pm 2.5$ & $1.07\pm 0.38$ \\
                  & G196.55+31.96 & $18.5\pm 3.5$ & $1.29\pm 0.64$ \\
G221+35 & G220.54+34.70 & $13.7\pm 1.9$ & $2.34\pm 0.44$ \\
G254+63 & G253.01+61.23 & $14.7\pm 3.5$ & $1.44\pm 0.86$ \\
G94-36    & G94.19-37.37 & $13.5\pm 1.3$ & $1.83\pm 0.28$ \\
          & G91.71-34.32 & $14.0\pm 1.4$ & $2.06\pm 0.29$ \\
          & G94.36-36.31 & $14.1\pm 1.6$ & $2.12\pm 0.40$ \\
          & G92.07-36.71 & $12.6\pm 1.2$ & $2.27\pm 0.34$ \\
          & G94.28-36.61 & $15.6\pm 2.4$ & $1.83\pm 0.45$ \\
          & G92.34-34.41 & $13.4\pm 1.8$ & $2.08\pm 0.49$ \\
          & G94.04-34.44 & $15.2\pm 2.5$ & $1.80\pm 0.52$ \\
          & G92.59-35.55 & $15.3\pm 2.5$ & $1.85\pm 0.59$ \\
          & G92.06-34.59 & $13.1\pm 3.5$ & $2.33\pm 1.06$ \\
          & G94.68-37.21 & $16.9\pm 2.8$ & $1.61\pm 0.52$ \\
          & G92.38-37.43 & $16.5\pm 2.4$ & $1.73\pm 0.42$ \\
          & G93.95-34.17 & $15.0\pm 3.0$ & $1.96\pm 0.74$ \\
          & G92.23-35.70 & $16.2\pm 2.9$ & $1.71\pm 0.52$ \\
G96-51  & G96.05-50.34 & $17.7\pm 1.8$ & $1.39\pm 0.23$ \\
          & G95.72-49.63 & $17.5\pm 2.5$ & $1.67\pm 0.46$ \\
G236+39 & G235.60+38.28 & $14.6\pm 1.9$ & $2.11\pm 0.46$ \\
          & G235.67+38.00 & $16.5\pm 2.2$ & $1.91\pm 0.42$ \\
\end{tabular}
\tablenotetext{a}{Designation within the Planck Galactic Cold Clump Catalog \citep{pgcc}}
\end{flushleft} 
\end{table}

\section{Results}

\subsection{Properties of the clouds}

{\bfc For each cloud, the dust temperature was measured by {\it Planck} and {\it IRAS}  using 
Equation~\ref{eq:inu}.}
In no case was one of the clouds of our sample warmer than its surroundings, indicating that
none of them are heated internally by stars.
Table~\ref{derivtab} lists the dust temperatures, $T_{\rm diff}$, in the diffuse medium around the clouds as well as the dust temperatures, $T_{\rm cloud}$, in the brightest parts of the clouds.
The dust temperatures in the clouds are typically 1 K lower than the surroundings, with temperature decrements into the clouds spanning 
0 to 1.6 K.

Of critical importance for this project is the manner in which the dust optical depth was derived from
the observed intensities. 
The dust column density is derived from multi-frequency measurements that cover the spectral energy distribution from
100 to 850 $\mu$m, spanning both sides of the peak and including sufficiently long wavelengths to measure the dust temperature
and wavelength-dependent emissivity. 
The emissivity index, $\beta$, was determined using maps degraded in resolution to $30^\prime$, in
order to improve sensitivity, while the optical depths were then calculated at the full $5'$ resolution. This will lead to systematic errors in locations where $\beta$ is spatially variable. In fact, the clouds in this
study tend to be local maxima in $\beta$. If $\beta$ were measured at full resolution, the maxima would likely become more pronounced. Because the dust temperature and $\beta$ are inversely correlated {\bfc  \citep{juvela13},} 
the dust temperatures in the clouds would likely decrease if $\beta$ were measured at full resolution.
For these reasons, we expect the cloud temperatures are likely to be even lower than the surroundings.
The amount of dust required to explain the observed emission increases if the dust temperature
is lower. 
Therefore, for the same observed emission from the clouds, we expect a full-resolution determination
of $\beta$ and $T_{\rm cloud}$ would yield dust optical depths {\it higher} than we report. 
The amount by which $\beta$ might increase is not known but is likely no more than 0.2 
{\bfc (based on the observed range of $\beta$ on large scales and not expecting drastic spectral variations
on angular scales between $5'$ and $30'$)}, with
implied systematic uncertainty of the cloud optical depths at the 30\% level.
We emphasize this point because we are measuring dust excess relative to gas in clouds, and
a full-resolution measurement would tend to make the dust excesses even more pronounced than
we measure.

To give some context for the clouds, Table~\ref{derivtab} lists the mass of each cloud, measured using the maps of $\tau(353\,{\rm GHz})$ and the standard
gas-to-dust mass ratio of 124 \citep{lidraine},
and the H~I mass, measured using the 21-cm line integral and assumed optically thin. Both mass determinations were made in 80$'$ radius regions (grown to 100$'$ for the largest cloud) centered on the infrared peak, with the unrelated emission in front of and behind the cloud estimated in an annulus surrounding it.
We assume the clouds are at a distance 100 pc, typical of the scale height for cold atomic gas \citep{dickeylockman}.
The clouds are typically 100 $M_\odot$.

\subsection{Comparison between dust and gas}

Figures~\ref{G254mainFigure} through~\ref{G96mainFigure} summarize the observations. The top panel of each Figure shows a color composite image combining the Arecibo and {\it Planck} data. The fields are filled with emission from the diffuse interstellar medium, with relatively bright features, which we noticed from the all-sky images as relatively isolated clouds, near the center. The lower panels show pixel-by-pixel comparisons of the images in the form of Hess diagrams, wherein we count the number of image pixels that fall within 
each of $300\times 300$ cells in the $N({\rm H I}), \tau({\rm 353\, GHz})$ plane.
For the lowest-column densities, which pervade much of the area of the images, and hence most of the points in the Hess diagrams, there is a fairly tight correlation between dust and gas. This correlation is evident by the relatively constant color of the spatial images, and the high density of points forming a red ridge in the Hess diagrams. 

For the distinct cloud-like features, the dust and gas trace similar structures but with distinctly different 
relative amounts.
For example, in Figure~\ref{G254mainFigure}, the ellipses labeled A, B, C show the spatial location (top panel) and the dust-gas correlation (bottom panel) of three features. 
The coherence of the features in the image {\it and} in the Hess diagram proves that the features have different relative amounts of dust and atomic gas.
It is evident that within the cloud-like features, there is relatively {\it more} dust per atomic gas than there is in the more diffuse medium. The individual clouds in the top panel correspond to `tracks' in the bottom panel due to their different dust-gas proportions.
In Figure~\ref{G108mainFigure}, the circles labeled A and B tell similar stories. {\bfc Because the `green' emission is associated with the
structure of the H~I cloud, it is almost certainly all interstellar. But because galaxies could also cause infrared excess (they gas being
at velocities outside the 21-cm survey), we investigated the relatively compact features in the ellipse labeled B individually.} The 
easternmost `green' feature in ellipse B is at 00$^{\rm h}$30$^{\rm m}$58.9$^{\rm s}$ +10$^\circ$46$'$30$''$, and there are no corresponding sources in the SIMBAD database. The peak just to its west is at 00$^{\rm h}$29$^{\rm m}$38.1$^{\rm s}$ +10$^\circ$47$'$36$''$ and has a corresponding {\it IRAS} point source catalog entry at 100 $\mu$m only with a flux density of 3 Jy. Inspection shows the source is
extended and there is no optical counterpart; the most likely explanation is that the {\it IRAS} and {\it Planck} emission are both interstellar. 

\subsection{Measuring the gas/dust ratio}

\begin{figure}[th]
\includegraphics[scale=.7]{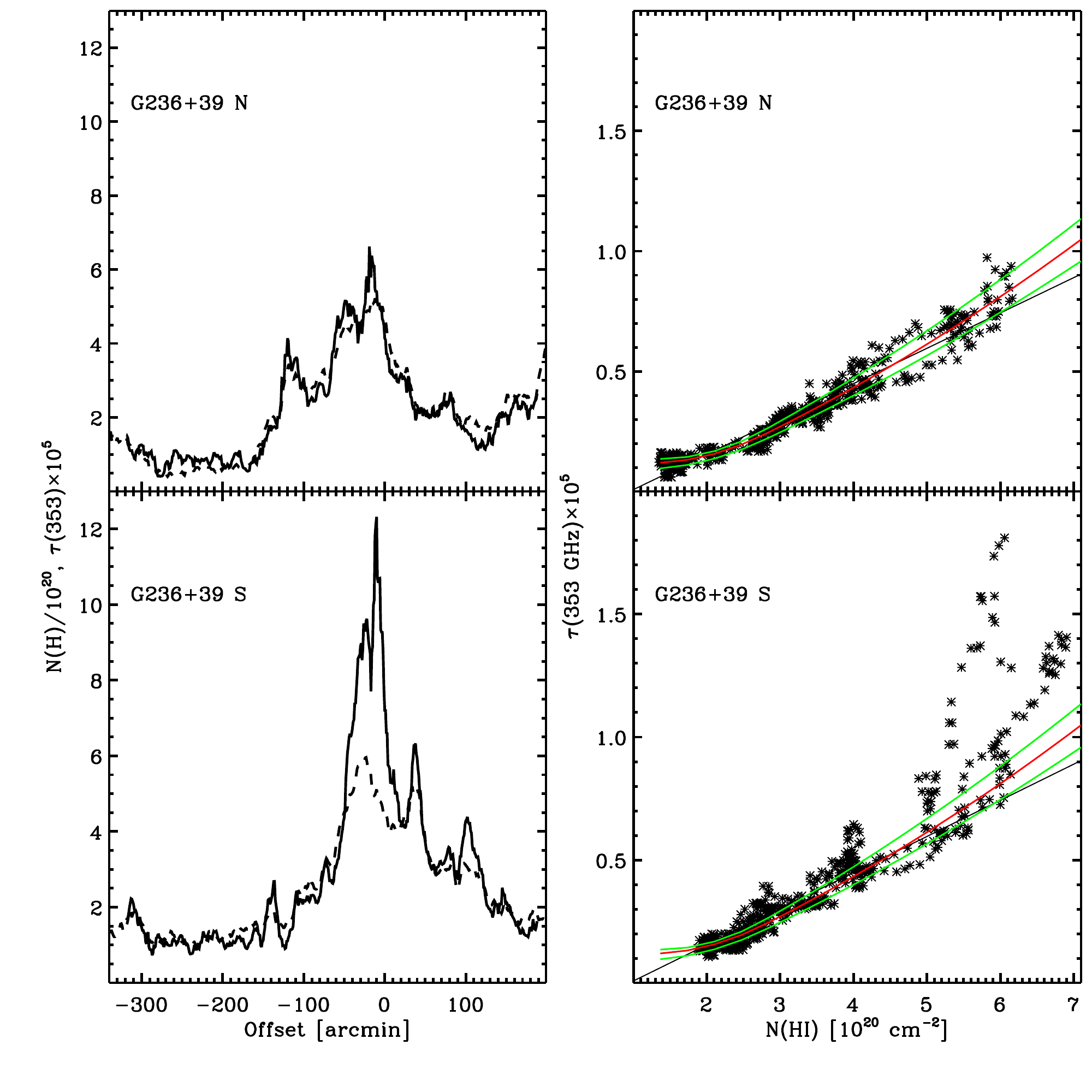}
\caption[g236cuts]{
Comparison of dust and gas in two horizontal slices through G236+39. 
{\bfc The upper panels are for a cut through the northern portion, G236+39 N, 
at constant declination (J2000)  +$02^\circ04'$, while
the lower panels are for a cut through the southern portion, G236+39 S, 
at constant declination +$00^\circ32'$. 
({\it left}) The H~I could density (dashed) and dust optical depth (thick solid) as a function of offset in right ascension along the slices. Notice how for G236+39 N (upper panel), the gas and dust follow each other closely. 
In contrast, notice how for G236+39 S (lower panel), there is a much sharper peak of the dust optical
depth than there is of the gas column density. 
({\it right}) Scatter plots of the dust optical depth versus H~I column density using the data from the left-hand panels. 
For G236+39N (upper panel), where the dust and gas are well correlated, a simple linear fit (thin solid line) fits the data showing the dust and atomic
gas are well mixed and linearly correlated. A red line shows a median B-spline fit, while the green lines surrounding it are the 25\% and 75\%
percentile splines.
For G236+39 S (lower panel), we over plot the same lines from G236+39 N, without scaling; this
shows that at low column densities the gas-to-dust ratio is the same as in G236+39 N,
while at higher column densities there are very large positive deviations indicating excess dust.}
\label{g236cuts}}
\end{figure}

The Figures of the clouds show that there are compact (`green') regions where there is material being traced by the
dust maps but not by the 21-cm maps. We can quantify the `excess' emission by scaling and subtracting the 21-cm map.
The measurement procedure is best summarized with an example.
Figure~\ref{g236cuts} shows slides through two components of the same cloud, G236+39.
Note how the northern component has a simple, linear trend of dust versus gas. The material along this slide
is most likely all atomic, and the 21-cm line is optically thin, so that both $\tau({\rm 353\, GHz})$ from {\it Planck} and
$N({\rm H~I})$ from Arecibo are linear tracers of column density of the same material. 
The slope of this line is the nominal dust optical depth per unit gas column density $\left<\tau/N({\rm H})\right>$. 
In contrast to the northern, purely atomic region, 
the cut through the southern component of the cloud shows a marked infrared excess at higher column densities.
The linear fit from the northern component is still valid for the outermost portions of the southern component, forming a lower envelope
at the higher column densities.
Now examine the spatial profiles of these clouds, in the right-hand panels of Figure~\ref{g236cuts}. While the dust and gas
are a very good match for the northern component, leaving little residual after subtracting the gas from the scaled dust map,
the southern component shows a significant positive excess, which is shown as the thin solid curve and is defined
as 
{\bfc 
\begin{equation}
N_{\rm ex} \equiv \frac{\tau({353\, GHz})}{\left<\tau/N({\rm H})\right>} - N({\rm H~I}). 
\end{equation}
In words, the infrared excess column density is the dust optical depth, scaled by the slope of the dust/gas correlation, then subtracting the
atomic gas. In Figure~\ref{g236cuts}, the excess can be directly measured from the right-hand panels by taking the positive
deviation of the dust optical depth above the straight line fitted at low column densities, then dividing that deviation by the slope of the
line fitted at low column densities.}
The dust opacity per unit gas column density and the infrared excess column density can now be 
 derived qualitatively, to interpret the color images of the clouds.

For each cloud, we identified the location of the infrared excess peak in the {\it Planck} $\tau({\rm 353\, GHz})$ map, and generated the H~I 21-cm line spectrum
from the Arecibo GALFA survey (subtracting a nearby off-cloud position to remove the large-scale H~I). 
In all cases, the velocity component in the H~I spectrum associated with the dust peak could be readily identified.
The central velocity and the full-width-at-half-maximum (FWHM) of these 21-cm line components is listed in Table~\ref{obstab}.
We then generated an H~I map covering the velocity channels centered on that line profile. These H~I maps in one velocity range had a much better
correspondence to the {\it Planck} images, for the cloud centered on the infrared excess peak. We then generated a one-dimensional profile through each
cloud, using a sliding box of width $1^\circ$ for each $1'$ pixel. The dust and velocity-selected H~I profiles showed peaks at the appropriate locations, and 
the slope of the gas-to-dust trend for these profiles was computed with a linear fit. 

The dust opacity per unit gas column density is directly related to the gas/dust mass ratio. The dust optical depth
\begin{equation}
\tau_\nu = \kappa_\nu N({\rm H}) \mu m_{\rm H}
\end{equation}
where $m_{\rm H}$ is the mass of a hydrogen atom, $\mu$ is the gas mass per unit H-atom taking into account other elements ($\mu=1.4$),
and $N({\rm H})$ is the hydrogen column density. Note that this is the {\it total} column density of H nuclei, including, if they exist and have
associated dust, molecular and ionized gas. The mass opacity coefficient, $\kappa_\nu$ depends only on the material properties
and size distribution of the dust grains. Using $\kappa({\rm 353\, GHz; 850\,}\mu{\rm m})=0.42$~cm$^2$~g$^{-1}$ from \citet{lidraine}, we can directly convert the observed dust opacity per unit gas column density into a gas/dust mass ratio:
\begin{equation}
\left[G/D\right]= 9.7\times 10^{-25} / \left[\frac{\tau({\rm 353\,GHz})}{N({\rm H~I})}\right]  
\label{eq:gdkappa}
\end{equation}
where the H~I column density is expressed in cm$^{-2}$.
The gas/dust mass ratios for the clouds are listed in Table~\ref{obstab} as
$\left[ G/D \right]_{\rm cloud}$.
To measure the gas/dust ratio for regions surrounding the clouds, we took the total H~I column density, minus the component centered on the cloud, and 
measured the gas/dust slope across the 64 deg$^{-2}$ region around each cloud, using a robust fit that excludes positive outliers.
These values are listed in Table~\ref{derivtab} as
$\left[ G/D \right]_{\rm diff}$. Please keep in mind that the $[G/D]$ values assume all the gas is atomic. We discuss possible
interpretations below, but a low value of $[G/D]$ generally means means
there is a deficiency in the inventory of gas.

It is evident that the gas/dust ratios are significantly higher in the diffuse medium than in the clouds. 
In the diffuse interstellar medium, the wide-field observations by {\it COBE} used in the \citet{lidraine} model
yield a gas/dust mass ratio of 124. This value is an ``upper envelope'' for the regions in the present Arecibo/{\it Planck}
study that was centered on infrared excess peaks. 
\def\extra{
The absolute value of the gas-to-dust ratio depends inversely on $\kappa$; if we assume that the places with dust temperatures 
greater than 19.5 K have the same grain properties as the diffuse ISM and gas is well traced by the 21-cm line emission, then
we would infer $\kappa$ is higher by a factor of $\sim 2$. For the remainder of this paper we will stay with the \citet{lidraine} value
of $\kappa$. The absolute value of $\kappa$ will have little effect on the results because we will primarily utilize the ratios of gast-to-dust ratio
to investigate changes of the ISM properties in the cloud cores.
}
The cloud cores all have gas/-to-dust mass ratios significantly
lower than the diffuse interstellar medium. Many of the regions surrounding the clouds have lower gas/dust ratio than the diffuse ISM,
suggesting that the same effect that causes the infrared excess peaks spreads over significant fractions of the 
regions studied. 

\clearpage

\section{Discussion}

\subsection{Trends of Dust/Gas}

\begin{figure}[th]
\includegraphics[scale=.8]{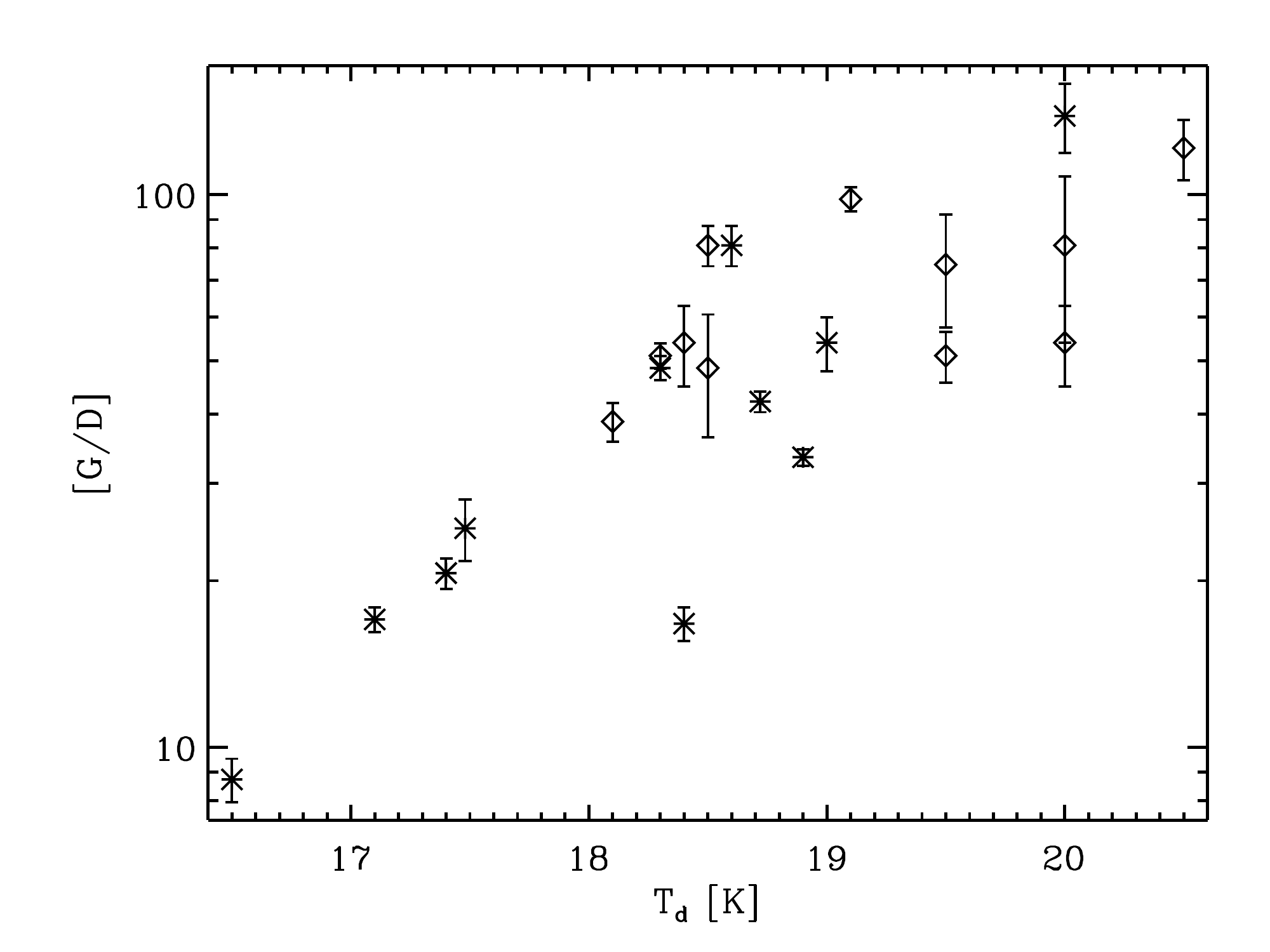}
\caption[cloudtab]{
The gas/dust mass ratio, measured from the optical depth at 353 GHz per unit H~I column density, versus the dust temperature.
The cloud peaks are shown as asterisks (*), and the diffuse regions around the clouds are shown as diamonds ($\diamond$).
\label{cloudtab}}
\end{figure}

The gas/dust ratio varies systematically with the temperature of the dust. 
Figure~\ref{cloudtab} shows the gas/dust ratio versus dust temperature, combining the observations toward the cloud peaks and the diffuse medium near the clouds. The diffuse medium dust temperatures were obtained from modes of the histograms of 
the $8^\circ\times 8^\circ$ regions surrounding the cloud. 
The clouds have {\it lower} gas/dust ratio compared to the interstellar gas-to-dust ratio of 124 \citep[][]{lidraine}; the cloud peaks tend to be lower
than the regions  surrounding them diffuse medium; and the colder clouds and diffuse regions have 
lower gas/dust than the warmer ones. The dynamic range of the trend is an order of magnitude.

The apparently excess dust is not an artifact of converting the observed infrared intensities into the dust optical depth.
The dynamic range of the dust temperature variations is
only 17--20 K, over which the predicted emission at 353 GHz changes by only 28\%, which is far smaller than the change in gas/dust. 
More likely, the
dust temperature in the cloud cores is {\it over}estimated because {\bfc of temperature gradients.
The dust temperatures in the cloud cores are lower  than that of their surroundings. If there is a radial temperature gradient into the
core, the central temperature is lower than the apparent temperature of dust as seen by us, because both the core and the warmer
dust surrounding it are present on the same line of sight.}
In this case, the optical depth is {\it under}estimated and the actual gas/dust in the clouds is actually even {\it lower} than inferred from the
observed line-of-sight averages. The decreased gas/dust ratio in the clouds is not due to the conversion from infrared intensity to dust column density,
and we turn to physical explanations for the effect.

Hypotheses for the observed trend are (1) the gas content neglects to include `dark' molecular gas 
(\S\ref{sec:darkgas});
(2) the amount of atomic gas in the clouds is underestimated because the 21-cm line is optically thick (\S\ref{sec:coldhi}); and
(3) the dust properties change such that the optical depth per unit dust mass is higher in colder clouds(\S\ref{sec:dustpropvar}). 
We discuss each of these hypotheses in turn in the
following subsections.

\subsection{Interpreting as `Dark Gas'\label{sec:darkgas}}

If we interpret the dust emission as a tracer of the total gas column density, using the correlation in the diffuse medium, then inverting Eq.~\ref{eq:gdkappa},
\begin{equation}
 N({\rm H})_{\rm dust} \equiv 1.03\times 10^{24} \tau({\rm 353\, GHz}) \langle G/D\rangle_{\rm diff},
\end{equation}
where $\langle G/D\rangle$ is the gas-to-dust
ratio measured measured in a reference region near
the cloud or taken from an interstellar average ($\langle G/D\rangle_{\rm ISM}=124$).
If the low values of gas/dust mass ratio are due to dust associated with some ``dark'' gas, then the column density of that gas
\begin{equation}
N({\rm H})_{\rm dark} = N({\rm H})_{\rm dust} - N({\rm H~I}).
\end{equation}
Note that this column density is the equivalent number of nucleons, as for atomic gas; if the gas were molecular, then $N_{\rm dark}=2N({\rm H}_2)$.
The factor by which the total gas column density exceeds the atomic gas is then
\begin{equation}
f_{\rm dark} \equiv \frac{N({\rm H})_{\rm tot}}{N({\rm H~I})} =  \frac{ [ G/D]_{\rm diff}}{\,\,\,[G/D]_{\rm cloud}},
\label{eq:fdark}
\end{equation}
measures the amount of `dark' gas with a unitless quantity that is directly (and linearly) related to observables. (Note that the dust properties needed for these
calculations, $\kappa$, cancel out in Eq.~\ref{eq:fdark} as long as they remain constant from cloud to more diffuse ISM.)
If $f_{\rm dark}=1$ then the cloud has the same gas/dust as the surrounding medium (and no `dark' gas is required),
while larger values indicate the ratio of total column density to that of atomic gas. 
The values of $f_{\rm dark}$ are summarized in Table~\ref{derivtab}.
For all clouds but one, the inferred amount of `dark' gas is larger than that of atomic gas, with total masses exceeding atomic gas by
factors 2--4. \footnote{In many cases, 
one should consider $f_{dark}$ calculated in this manner as lower limits, because the `dark' gas may extend far  from the cloud and
into the region where $[G/D]_{\rm diff}$ is calculated. We use the conservative values in the remainder of the paper, but note for now that if instead we
used the diffuse ISM gas/dust ratio value, $\langle G/D\rangle_{\rm ISM}$ as the reference, the values of $f_{\rm dark}$ would be typically 3--8.
}
The fraction of the total amount of gas that is `dark' is $1-f_{\rm dark}^{-1}$, with typical values 50\% to 75\%.
In smaller patches comparable to the $5'$ beam size, the relative amount is significantly higher, reaching up to 6, meaning dark gas comprises 83\% of the total. 
The ratio of `dark' mass integrated over the clouds can be estimated similarly, though without taking into account the gas-to-dust ratio of the environment,
using the cloud mass determined from the dust and H~I maps; values of $M_{\rm cloud}/M_{\rm H~I}$ are similar to $f_{dark}$ in the range 1--5.

The most likely candidate for `dark' gas is H$_2$. 
Molecular hydrogen forms readily in interstellar conditions and would dominate the ISM if it were not dissociated by ultraviolet photons from the diffuse interstellar radiation field. 
In this hypothesis, $N({\rm H}_2)=\frac{1}{2} N_{\rm dark}$ measures the column density of molecular gas. This hypothesis can be tested in many ways: comparison to existing tracers of molecular gas, comparison to results of absorption-line observations of molecular gas along essentially random lines of sight (toward ultraviolet sources), and searches for new tracers of molecular gas. The last idea is the subject of follow-up observations we are pursuing, using OH absorption and [C II] emission. We address each of the other two tests now. 

\subsubsection{Comparison to CO}

For high galactic latitude regions, there is no sensitive all-sky survey of the best mapping tracer of molecular gas, which is the millimeter-wave CO(1--0) line. However it was found that the {\it Planck} 100 GHz band contained significant emission from the CO(1--0) line that could be separated from the continuum emission using a multi-frequency analysis \citep{planck2013-p03a}. The Type 3 CO map from that paper has a resolution of $5.5'$ and rms fluctuations of the (mostly empty) map near one of our selected clouds are 0.5 K~km~s$^{-1}$ at this angular scale.  
For molecular clouds in the galactic plane, the typical ratio $X({\rm CO})\equiv N({\rm H}_2)/W({\rm CO})=2\times 10^{20}$ cm$^{-2}$~K$^{-1}$~km$^{-1}$~s, so the {\it Planck} CO map would be sensitive to $N({\rm H}_2)\sim 10^{20}$~cm$^{-2}$ if that same ratio applied.
For comparison, the map of $N_{\rm dark}$ has an rms at $6'$ angular scale of $0.4\times 10^{20}$ cm$^{-2}$, which would correspond to molecular column densities of $0.2\times 10^{20}$~cm$^{-2}$. Thus the $5\sigma$ clouds in the 
{\it Planck} maps of $N_{\rm dark}$ should appear in the CO map, if they have the same scaling as galactic plane molecular
clouds. 

\begin{figure}[h]
\includegraphics[scale=.7]{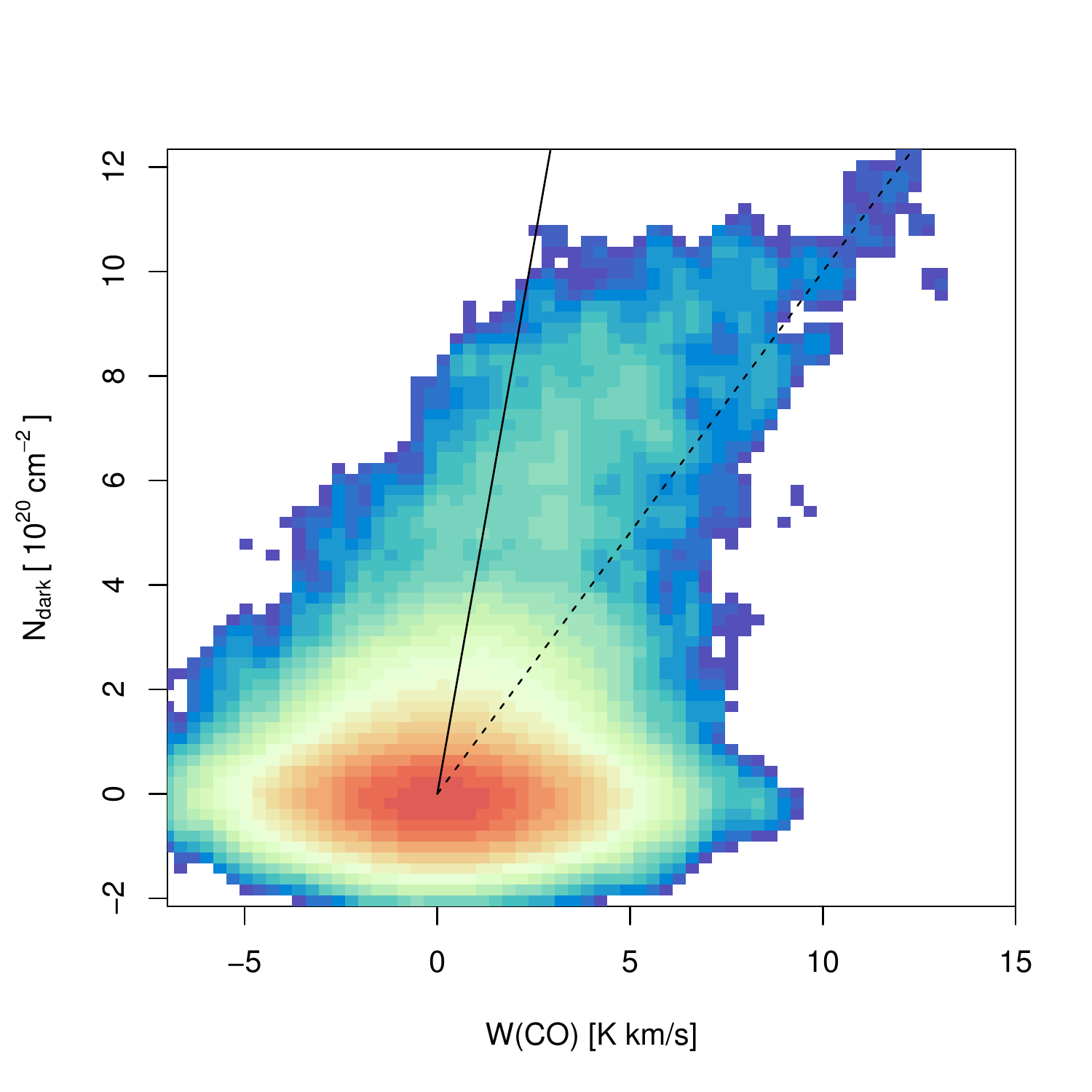}
\caption[coplotnew]{
The `dark' nucleon column density versus the CO line integral for the cloud G108-53. They color image is the Hess diagram showing the density
of pixels at each value of $N_{\rm dark}$ and $W({\rm CO})$, generated with
a gaussian kernel with width 0.5 in each axis (in the units shown). 
The straight dashed line is the slope that would be expected based on the galactic plane value of $X_{\rm CO}=2.1$, while the 
dotted line is a rough lower bound at $N_{\rm dark}/W({\rm CO})\simeq 1$, which  corresponds
to $X_{\rm CO}=0.5$.
\label{coplotnew}}
\end{figure}

\begin{table}[h]
\caption[]{``X-factor'' for Infrared excess clouds}\label{corattab} 
\begin{flushleft} 
\begin{tabular}{lcccc}
\hline\hline
Name &  $\langle \frac{1}{2} N_{\rm dark}/W({\rm CO})\rangle$\tnm{a} & $ \frac{1}{2} N_{\rm dark}/W({\rm CO})\tnm{b}$ \\
\hline
G94-36  & $0.45\pm 0.05$ & $0.34\pm 0.05$\\
G96-51  & $2.3\pm 0.4$ & $2.8\pm 0.3$\\
G104-39 & $0.60\pm 0.1$ & $0.60\pm 0.1$ \\
G108-53 & $0.75\pm 0.1$ & $0.65\pm 0.1$ \\
G138-52 & $1.1\pm 0.3$ & $0.90\pm 0.1$ \\
G198+32 & $>4$ & $>5$\\
G221+35 & $>3$ & $>4$ \\
G236+39 & $1.2\pm 0.2$ & $1.4\pm 0.1$\\
G254+63 &  $>10$ & $>20$\\
\end{tabular}
\tablenotetext{a}{half the slope of the infrared excess column density, $N_{\rm dark}$, versus CO line integral, $W({\rm CO})$, from the Hess diagram}
\tablenotetext{b}{half the ratio of $N_{\rm dark}$ over CO line integral $W({\rm CO})$ at the cloud peak}
\end{flushleft} 
\end{table}  

The {\it Planck} CO map for some clouds shows a good correspondence with the infrared excess, in particular
G94-36, G104-39, and G108-53; others show a CO detection over a limited portion of the infrared excess region, 
such as G96-51, G138-52, and G236+39; while the rest show no CO detection.
Figure~\ref{coplotnew} shows a pixel-by-pixel comparison of the $W({\rm CO})$ and $N_{\rm dark}$ for G108-53.
The slope expected for the $X$-factor of the galactic plane molecular clouds forms a rough upper bound of the points,
while a lower bound is formed by a line with a 4 times lower effected $X$-factor.
The CO and `dark' gas maps agree in detail for this cloud, and after subtracting the 
CO image scaled by the dashed line, the prominent infrared cloud that is green in Figure~\ref{G108mainFigure} is removed
cleanly. To quantify the correlation between CO and infrared excess, we made Hess diagrams like Figure~\ref{coplotnew}
for each cloud and estimated the slope $\frac{1}{2} N_{\rm dark}/W({\rm CO})$. (The factor of $\frac{1}{2}$ is needed because we
calculate the dark gas column density in units of H nuclei, while the $X$-factor is for H$_2$ molecules.)
For many clouds, only lower limits were possible due to non-detection of CO.
Then we identified the peaks in the infrared excess and the CO map (when present in the latter) and measured the ratio 
directly on the peaks. 
Again, some ratios are lower limits due to lack of detected CO, and we use 95\% confidence ($2\sigma$) limits.
Table~\ref{corattab} summarizes the results.

For the brightest infrared-excess clouds with clearly associated CO peaks, 
the values of $X({\rm CO})$ we find  are about a factor of 2 lower than the galactic plane value. 
{\bfc For G94-36 there is a linear correlation between $N_{\rm dark}$ and $W({\rm CO})$, yielding an accurate
measure of $X({\rm CO})$. 
Figure~\ref{coplotnew} shows for G108-53 a range of apparent slopes between $N_{\rm dark}$ and $W({\rm CO})$, possible curving from a 
high slope in faint regions to a lower slope in bright ones
(though this cannot be quantified accurately due to the low signal levels). }
For other clouds, the lower limits on $X({\rm CO})$  exclude the galactic plane value \citep[cf.][]{xfactorreview} 
{\bfc as well as the value for the brighter infrared excess clouds from this paper}.
The high $X({\rm CO})$ for those clouds does not rule out the presence of H$_2$ with faint CO. 
The linear scaling between CO brightness and H$_2$ column density only applies to molecular clouds that are gravitationally
bound \citep{xfactorreview}. If the CO line is optically thin or sub thermally excited,  $X({\rm CO})$ can be high even
if CO is abundant, so we cannot rely upon the CO(1--0) emission as a tracer in diffuse regions.

\def\extra{
\subsubsection{Comparison to other tracers of molecular gas}

* UV FUSE results
* future test with other molecular tracers including our OH observation with Arecibo
}

\subsubsection{Trend of dark gas column density with dust temperature}

\begin{figure}[th]
\includegraphics[scale=.7]{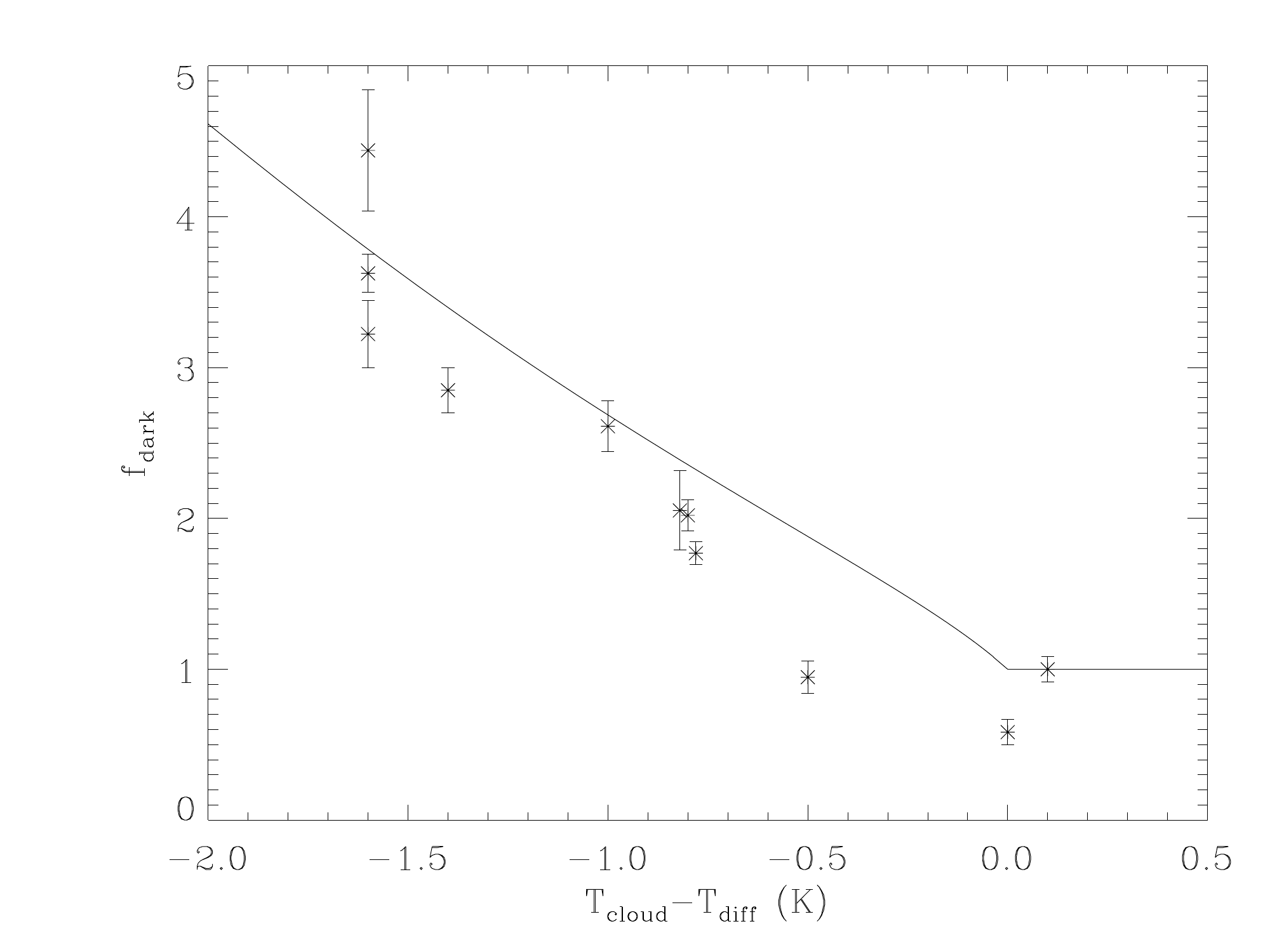}
\caption[cloudtd]{
The ratio of total column density to atomic column density, $f_{\rm dark}$, for each cloud versus the dust temperature difference
between the interior of the cloud and its surroundings. 
\label{cloudtd}}
\end{figure}

The inferred amount of `dark gas' depends very cleanly on the {\it difference} in dust temperature between the cloud and its surroundings. 
Figure~\ref{cloudtd} shows inferred amount of `dark gas'  increases monotonically with the dust temperature difference between the cloud interior and surroundings. This trend has a relatively straightforward interpretation. Within cold cloud interiors, molecules form and the H~I is no longer tracing the total column density. In the context of the `dark gas' hypothesis, the cold regions are where the dust is shielded from
the radiation field, and those are the same regions where H$_2$ self-shields.
As the amount of self-shielding increases the dust temperature decreases, and the molecular hydrogen formation rate surpasses the
dissociation rate allowing hydrogen to become largely molecular.
The regions with little self-shielding have dust temperature as warm as the widespread ISM and the gas remains photo dissociated.

The trend between dark gas and dust temperature can be reproduced by a simple model. The abundance of H$_2$ depends on the
abundance of H~I from which it forms, as well as the dissociation rate; for a clod with uniform density 100 cm$^{-3}$ and gas temperature 80 K,
\begin{equation}
\frac{2N({\rm H}_2)}{N({\rm H~I})} = 0.8 \chi^{-2} N({\rm H~I}),
\label{eq:nhtwo}
\end{equation}
where $\chi$ is the amount of radiation  relative to the diffuse interstellar radiation field \citep[eq. 3]{reach94}.
The energy emitted (or absorbed) by dust depends on $T^{4+\beta}$, so the dust temperature difference is related to the radiation field by
\begin{equation}
T=T_0 \chi^{\frac{1}{4+\beta}}
\label{eq:tt0}
\end{equation}
where $T_0$ is the dust temperature at the surface of the cloud. 
The attenuation of the radiation field is roughly approximated as
\begin{equation}
\chi=e^{-2 A_V/1.086}
\end{equation}
where the visible-light extinction is
\begin{equation}
A_V=\left[ N({\rm H~I}) + 2 N({\rm H}_2) \right] / 1.9\times 10^{21} {\rm~cm}^{-2}
\label{eq:avnh}
\end{equation}
\citep{savagemathis}.
The factor of 2 in the equation for $\chi$ is to account for the absorption of dust-heating photons being at ultraviolet rather than visible 
wavelengths. 
{\bfc Note that the gas-to-dust ratio in Equation~\ref{eq:avnh} is consistent with the assumed value for the diffuse interstellar medium
by construction because it was part of the dust model from which obtained $\kappa$.}
The prediction of this simple model is shown in Figure~\ref{cloudtd} and agrees quite well with the observed data. Thus the idea
of the infrared excess being due to dust associated with `dark' molecular gas can match the observations quantitatively.

\subsection{Interpreting as optically thick H~I\label{sec:coldhi}}

We have assumed throughout this paper that the integral over the 21-cm line brightness traces the column density of the atomic gas linearly.
In fact, the 21-cm line is prone to optical depth effects over the column densities of interest.
The under prediction of the H~I column density by the optically thick 21-cm line could account for the apparent excess dust column density
in interstellar clouds, as was recently suggested by e.g. \citet{fukui15}.
The optical depth 
\begin{equation}
\tau(v) = N(v) / C T_s,
\end{equation}
where $v$ is the velocity, $N(v)$ is the column density per unit velocity at velocity $v$, $C=1.823\times 10^{18}$~cm$^{-2}$~K$^{-1}$~km$^{-1}$~s,
and the line brightness 
\begin{equation}
T_B(v) = T_s [1-e^{-\tau(v)}],
\end{equation}
where $T_s$ is the excitation (`spin')  temperature \citep{kulkarniheiles}. 
The column density in the optically thin limit \[ N_H^* \equiv C \int T_B(v) dv \]
while the actual column density \[ N_H \equiv \int N(v) dV = C T_s \int \tau(v) dv. \]
To calculate the total column density, one must correct the column density in the optically thin limit by a factor
\begin{equation}
 \frac{N_H}{N_H^*} = \left< \frac{\tau}{1-e^{-\tau}} \right>  \equiv \frac{\int T_B(v) \frac{\tau(v)}{1-e^{-\tau(v)}}dv}{\int T_B(v) dv}. 
\label{eq:nhnhraw}
\end{equation}
For a single gaussian component with peak optical depth $\tau_0$, we find a good fit
\begin{equation}
 \frac{N_H}{N_H^*} \simeq \left[ \frac{\tau_0}{1-e^{-\tau_0}} \right]^{0.75} 
\label{eq:tauapprox}
\end{equation}
 over a wide range of optical depths ($0.01 < \tau_0 < 20$) and independent of spin temperature.

An advantage of studying isolated clouds at high galactic latitude is that the observations are dominated
by  a single velocity component of 21-cm emission and a single spatial
feature in both 21-cm and infrared emission.
If the clouds we are observing are entirely atomic gas, and $N_{\rm dust}$, as in the previous section, measures the total column density,
then the ratio
\begin{equation}
 \frac{N_{\rm dust}}{N_H^*} = f_{\rm dark} 
\end{equation}
which can be combined with Equation~\ref{eq:tauapprox} to solve for the optical depth of the 21-cm line.
Taking the observed $f_{\rm dark}$ for our clouds from Table~\ref{derivtab}, the implied 21-cm line optical depths of the clouds are of order unity,
with typical values $\tau_0\sim 3$. If the 21-cm lines are indeed this optically thick for the majority of the clouds, then observations of radio
sources will show deep absorption features. The results of the Millennium H~I survey \citep{heilestroland03} indicate that 
optical depths of the 21-cm line are typically small, but a pencil-beam survey cannot be used to address the possibility of regions
of high 21-cm optical depth that coincide with where we observe high gas/dust ratios. 
The possibility that the apparently low gas/dust ratio in isolated clouds could be due to the H~I being optically thick 
can be followed up with
future observations of radio sources situated behind the clouds, and will be the subject of future observational follow-up.

In a recent paper, \citet{fukui15} suggest that the excess infrared emission often referred to as `dark gas' is optically thick H~I, which is the same hypothesis we 
explore in this section of this paper. \citet{fukui15} measure the temperature variation of dust using {\it Planck} and find it is correlated with the apparent infrared
excess, then use this empirical correlation to determine the total column density under the assumption the extra infrared emission is from optically
thick H~I. Shielding of the radiation field from the interior of clouds is a straightforward explanation that must be contributing to the apparent
trends. The simple model we presented in the previous section (Eq.~\ref{eq:nhtwo}--\ref{eq:tt0}) and illustrated in Figure~\ref{cloudtd} seem to match 
the observations for the clouds in our sample well. The lack of widespread optically thick H~I from 21-cm absorption studies further argues
against significant an amount  cold H~I adequate to explain the range of gas/dust variations that are observed.
On a technical note, our equations above (Eq.~\ref{eq:nhnhraw}--\ref{eq:tauapprox}) 
are similar to theirs; however, they authors did not take into account the line profile and defined a
single optical depth for all velocities along each line of sight. 
This fictitious optical depth , $\tau^*$, can be solved from the line profile and Equation~\ref{eq:nhnhraw}:
\[ \frac{\tau^*}{1-e^{-\tau^*}} = \left< \frac{\tau}{1-e^{-\tau}} \right>. \]
\citet{fukui15} use the fictitious $\tau^*$ together with a fictitious line-width $\Delta V$ (the ratio of the line integratal to the 
peak brightness temperature), and a single spin temperature $T_s$,
as intermediaries to convert the observed line brightness into a total column density.
None of $\tau^*$, $\Delta V$, or $T_s$ apply on typical lines of sight where the WNM contains about half the column density
and has a distinctly different profile than the CNM. 
The Millennium Survey results showed that randomly selected lines of sight toward extragalactic sources have multiple
velocity components of varying widths and temperatures \citep{heilestroland03}.
For the simplest case of a single CNM line on the line of sight, like for the clouds in the present paper, the concepts 
are appropriate, and the more detailed solution finds that
\citet{fukui15} would systematically overestimate the true column density by an amount that varies from
0 at low optical depth, to 15\% at $\tau_0=1$, to 50\% at $\tau_0=5$, where $\tau_0$ is the peak optical depth of the 
line profile.

Our results being produced solely by optically thick 21-cm line is
highly unlikely because it requires typical optical depths $\ge 3$. Such
high optical depths are rare: \citet{heilestroland03} surveyed 79
21-cm line absorption spectra and, apart from two positions in the
Galactic plane, only one exhibited an optical depth exceeding 3. Also,
\citet{stanimirovic14}
investigated whether the
optical depth effect was significant in their inferring total H$_2$
content by comparing dust IR emission and 21-cm line emission in the
Perseus molecular cloud by measuring 26 21-cm absorption line spectra in
the immediate vicinity, and concluded that the optical depth effect was
essentially negligible.

\subsection{Interpreting as dust property variations\label{sec:dustpropvar}}

The apparently-low gas-to-dust ratios observed toward the isolated clouds in this study could be due to the amount of dust being overestimated, which could happen if the dust properties are different in the clouds than in the diffuse interstellar medium. Recall the 
gas-to-dust ratios are calculated assuming a value of the dust opacity, $\kappa$, which we take from \citet{lidraine}. 
To explain a gas-to-dust ratio lower than the diffuse ISM by a factor of 3, we require that $\kappa$ is higher by a factor of 3.
Variations of this amount were indeed inferred from the first sub millimeter measurements of a diffuse interstellar cloud and comparison
to extinction from star counts \citep{bernard99,stepnik03}.
A recent  comparison of {\it Herschel} emission to near-infrared extinction for cold cores shows the ratio of 250 $\mu$m to 1.2 $\mu$m optical
depth is higher than in the diffuse ISM by a factor of 3 \citep{juvela15};
this is a `dust to dust' comparison unaffected by `dark' gas.
The earlier submm and {\it Planck} results both indicate higher $\kappa$, in the same direction and approximate amount required to explain the 
gas-to-dust variations derived in the present study.
Observations of gas and dust in circumstellar disks yields $\kappa$ about a factor of 10 lower \citep{beckwith90}; such 
a low $\kappa$, if applied to the ISM, would yield gas-to-dust ratios 10 times higher, which goes in the opposite direction from what is observed
in the cloud cores and is inconsistent with the observed depletions of metals in the diffuse ISM.

Because $\kappa$ is the ratio of the extinction cross section to the dust mass, it is sensitive to the size distribution of the 
large grains. 
For a compact (solid) particle, $\kappa$ is inversely proportional to the grain size, so grain growth would decrease $\kappa$.
But for coagulation of particles into fluffy aggregates, $\kappa$ apparently increases by a factor of 2--3
for silicate grains in the size range relevant to the ISM \citep{kohler12}.
Variations in the dust properties are
important to understand, not only because they may affect the inference of `dark gas' presence,
but also because of the role played by dust grains in the physics and chemistry of the interstellar medium,
as the primary agents for gas heating (through the photoelectric effect) and molecule formation (which occurs primarily on
grain surfaces).

\def\extra{

\begin{figure}[th]
\includegraphics[scale=.8]{cloudbeta}
\caption[cloudbeta]{
The emissivity index of the dust ($\beta$) versus the dust temperature ($T_d$). Values with error bars are the center of each of the clouds in our
sample. The color image is a Hess diagram showing frequency of each value of $T_d,\beta$ within the 
combined $8^\circ\times 8^\circ$ regions centered on the clouds in our sample. The cloud cores, where the enhanced apparently gas-to-dust 
ratios exist, tend to be at the edges of the distributions of  $T_d,\beta$ for their surrounding fields, which shows that they tend to have
dust properties distinct from the surrounding ISM. 
The large, filled circle is the diffuse ISM value derived from the faintest 10\% of the entire sky; this diffuse value is at the high-$T_d$, low-$\beta$ 
edge of the distribution for the cloud regions.  The two clouds most clearly deviating from the 
$\beta$-$T_d$ trend are labeled.
\label{cloudbeta}}
\end{figure}

The emissivity, $\kappa$, is challenging to measure because it requires comparison between very different tracers of material.
The spectral index, $\beta$, of the variation of dust emissivity with frequency can be measured  with  far-infrared
observations alone, and is a different way to find evidence for dust property variations.
p Note that the dust temperature and
emissivity index depend critically on the zero-points and band-to-band calibration, which took some time to be carefully measured by
the {\it Planck} team and which differ from earlier results.
Figure~\ref{cloudbeta} shows the emissivity index and dust temperature for the clouds and, for comparison, the diffuse interstellar medium.
The overall tendency for higher $\beta$ at lower dust temperature is evident, though the scatter is significant. Examining individual clouds
in detail, the $\beta$ values are not noisy and the differences from cloud to cloud are real. 
The trend is in the same direction as was observed over a wider range of ISM conditions by \citet{dupac03}.

At wavelengths $>100$ $\mu$m, the grains are all in the small particle limit for
physical optics, and the changes in $\beta$ should primarily be due to changes in the grain composition \citep{coupeaud11}.
}

Changes in dust composition are constrained by the availability of elements.
Many heavy elements, including for example Fe and Si, are almost completely locked in grains \citep{dwek98}, so 
unless the regions with different dust properties are enriched in heavy elements, any new grain construction would need to be done 
without the refractory material already depleted onto the grains. We have found no evidence of central sources in the clouds that
could be responsible for heavy element enrichment. So for now we assume that regular cosmic abundances apply, and therefore
that the changes in properties may be related to the grain's geometry, degree of amorphization, or surface properties (like mantles).
The carbonaceous and silicate grains may combine to form larger grains, for example, which may increase $\kappa$. 
\def\extra{
This effect, if derived from physical optics alone,
 is really only expected to be
noticeable when the grain size approaches the observed wavelength; specifically when $2\pi a/\lambda$ approaches unity, where $a$
is the radius and $\lambda$ the wavelength. The {\it Planck} observations that determine $\beta$ extend to over 500 $\mu$m, which would
mean grain growth to radii of 80 $\mu$m, or three orders of magnitude from the size of `large' grains that produce most of the far-infrared
emission from the ISM \citep{lidraine}. Therefore the grains are not growing to size comparable to wavelength.
Instead, it is possible that the grains in dark clouds are changing in physical properties that affect the far-infrared cooling modes.
}
The wavelength-dependent emissivity may change as a result of any surface compositional change to the grains.
This could occur when small grains cover the surfaces of the large grains.
In this case it is not the `growth' in size that matters but rather the change in chemical bonds affecting the 
manner in which energy is absorbed by and emitted from the surface.
Carbonaceous material may evolve due to prolonged exposure to the interstellar radiation field, leading to a darkening over time
\citep{cecchi10,jones13}.
\rm The nature of the silicate material may also be evolving in dark clouds.
In order to explain the large values of $f_ {\rm dark}$ by increased
grain emission, one would have to invoke the quite unlikely scenario of
a greatly enhanced (factor $\sim 3$) metal abundance, all of which resides in
the dust; this seems unlikely.


\def\extra{
Thus dust property variations could be the explanation of the dust-to-gas variations.
Variations in the dust properties in interstellar clouds have implications for the thermodynamics of the ISM. 
Photoelectric heating from grain surfaces determines the temperature of interstellar gas. The photoelectric heating
is primarily from the smaller grains, so if those grains are removed, the gas temperature may be significantly decreased.
Molecule formation occurs on grain surfaces and depends on the temperature of the grains \citep{cazaux04}.
For these reasons, the changes in dust properties, even if subtle, may be linked to and indeed the trigger for
the changes in interstellar cloud properties discussed throughout this paper: altered grains could lead to colder atomic gas 
(which is more optically thick in the 21-cm line) and more molecular gas (which forms more readily on cold surfaces).
}


\clearpage

\section{Conclusions}

We find strong variations in the ratio of gas to dust mass both within interstellar clouds and from cloud to cloud. 
Our work complements all-sky studies, which use large amounts of data but, when averaging together, give a misleading impression of uniformity, missing the
true diversity of interstellar clouds.
The three hypotheses that we present as potential explanations for the variations all have implications for the interstellar medium, and they also have observational tests that can be performed.
The first two hypotheses are based on the idea that the gas column density may have been underestimated, while the true gas-to-dust mass ratio is constant everywhere. 
There would need to be extra gas, commonly referred to now as `dark gas' following \citet{grenier05}
not traced linearly by the 21-cm line emission.

The possibility of molecular gas (first hypothesis) would not be a surprise given its prevalence in the diffuse interstellar medium as measured
by UV absorption on random lines of sight \citep{radford02} and the widespread usage of dust images to trace the total gas column density even when
in excess of the atomic gas column density \citep{planckpaperondarkgas}.
For the brightest clouds in our sample, CO is detected and the H$_2$/CO ratio appears `normal' compared
to galactic plane clouds. But for many clouds, and for all cloud peripheries, there is no CO and the upper limits show that any H$_2$ present is 
far in excess of that expected based on the CO emission. The implication would be that the clouds in this sample are in fact {\it molecular clouds}, despite being 
translucent and tenuous. The clouds are {\it not} gravitationally bound, with gravitational forces being exceeded by the combined influence of both thermal 
and non-thermal (turbulent) motions by factors of $10^2$ to $10^3$. Observational tests can confirm or deny the presence of molecules, such as OH (which is
readily formed wherever H$_2$ is present) in the clouds at
abundance levels sufficient to explain the apparent  `dark gas' as H$_2$.

The second hypothesis is that the 21-cm line is optically thick, which is expected if the atomic gas is cold and has been proposed as an explanation for the 
apparent `dark gas'. We find that high optical depths are required, so that the 21-cm line is optically thick over entire interstellar clouds. The implications of this hypothesis are that 21-cm line observations are poor tracers of interstellar column density. The gas is likely to be cold and have optical depths $>3$ in the 21-cm line over significant areas. This hypothesis can be tested with absorption line measurements toward radio sources.  

The third hypothesis is that the dust grain properties vary, in which case there may be no `dark gas' at all, but instead a new physical process that changes grain properties.
If the grain properties evolve in translucent clouds, then the building blocks of molecular clouds also begin with these modified grain properties. Given the constraints of cosmic
abundances and the high efficiency with which dust absorbs starlight already, changes in dust grain properties are unlikely to explain the observed
gas-to-dust variations, but this hypothesis deserves to be explored and compared to independent observational data.

\acknowledgements  
The Galactic Arecibo L-Band Feed Array HI (GALFA-HI) Survey data set
was obtained with the Arecibo L-band Feed Array (ALFA) on the Arecibo
305m telescope.  
The Arecibo
Observatory is operated by SRI International under a cooperative
agreement with the National Science Foundation (AST-1100968), and in
alliance with Ana G. Mndez-Universidad Metropolitana, and the
Universities Space Research Association.

Facilities: \facility{Planck}, 
\facility{Arecibo},
\facility{IRAS}

\bibliographystyle{apj}
\bibliography{wtrbib,Planck_bib}

\end{document}